\documentclass[11pt]{article}
\usepackage[utf8]{inputenc}
\usepackage[margin=1in,left=1in]{geometry}
\usepackage{setspace}
\usepackage{graphicx}
\usepackage{amssymb, amsmath, amsthm, graphics}
\usepackage{mathabx,epsfig}
\usepackage[mathscr]{euscript}
\usepackage{tikz}
\usepackage{tikz-cd}
\usepackage{mwe}
\usepackage{pigpen}
\usepackage{bm}
\usepackage{mathtools}
\let\underbrace\LaTeXunderbrace    
\usetikzlibrary{calc}
\usetikzlibrary{matrix}
\usepackage{hyperref}

\usepackage{enumitem}

\usepackage{verbatim}
\usepackage{smartdiagram}

\usepackage{latexsym}
\usepackage[all]{xy}
\usepackage{color}

\usepackage{bbold}

\usepackage{tikz-cd}
\usetikzlibrary{positioning}
\usepackage{euscript}
\usepackage{multirow}
\usepackage{etex, pictexwd,dcpic}
\usetikzlibrary{positioning}
\usetikzlibrary{shapes.geometric}
\usetikzlibrary{shapes.misc}
\usetikzlibrary{calc}
\usetikzlibrary{positioning}

\usepackage{pgflibraryarrows}
\usepackage{pgflibrarysnakes}
\usepackage{tikz-3dplot}

\usetikzlibrary{trees} 
\usetikzlibrary[trees] 

\newtheorem{theorem}{Theorem}
\newtheorem{definition}[theorem]{Definition}
\newtheorem{lemma}[theorem]{Lemma}
\newtheorem{corollary}[theorem]{Corollary}
\newtheorem{proposition}[theorem]{Proposition}

\newtheorem{remark}{Remark}

\newtheorem{example}{Example}

\numberwithin{equation}{section}

\renewcommand{\(}{\begin{equation*}}
\renewcommand{\)}{\end{equation*}}
\newcommand{\bea}{\begin{eqnarray*}}
\newcommand{\eea}{\end{eqnarray*}}

\def\endofproof {\hfill{$\Box$}\\}

\newcommand{\beq}{\begin{equation}}
\newcommand{\eeq}{\end{equation}}

\newcommand{\into}{\hookrightarrow}

\newcommand{\op}[1]{\ensuremath{\operatorname{#1}}}

\newcommand{\theproof}{\noindent {\bf Proof.\ }}

\numberwithin{equation}{section}

\renewcommand{\(}{\begin{equation}}
\renewcommand{\)}{\end{equation}}

\def\ch{{\rm  ch}}

\def\1{{\bf 1}}

\def\<{\langle}
\def\>{\rangle}

\numberwithin{equation}{section}


\newcommand{\R}{\ensuremath{\mathbb R}}
\newcommand{\RR}{\ensuremath{\mathbb R}}
\newcommand{\NN}{\ensuremath{\mathbb N}}

\newcommand{\ZZ}{\ensuremath{\mathbb Z}}
\newcommand{\Z}{\ensuremath{\mathbb Z}} 
\newcommand{\QQ}{\ensuremath{\mathbb Q}}

\newcommand{\BB}{\ensuremath{\mathbf B}}

\newcommand{\CC}{\ensuremath{\mathbb C}}

\newcommand{\map}{\mathrm{Map}}

\begin{document}

\title{
Ramond-Ramond fields and twisted differential K-theory}

 \author{
 Daniel Grady and Hisham Sati\\
  }

\maketitle

\begin{abstract} 
We provide a systematic approach to describing the Ramond-Ramond (RR) fields 
as elements in twisted differential K-theory. This builds on a series of constructions
by the authors on geometric and computational aspects of twisted differential K-theory,
which to a large extent were originally motivated by this problem. In addition to providing 
a new conceptual framework and a mathematically solid setting,  
this allows us to uncover interesting and novel effects. 
Explicitly, we use our recently constructed Atiyah-Hirzebruch spectral sequence (AHSS) 
for twisted differential  K-theory to characterize the RR fields and their quantization,
which involves interesting interplay between geometric and topological data. 
We illustrate this with the examples of spheres, tori, and Calabi-Yau threefolds. 

 \end{abstract}

 \tableofcontents
 
 \newpage
\section{Introduction}

The goal of this paper is to combine proposals about the Ramond-Ramond (RR) fields 
in type II string theory,  going back to \cite{MW}\cite{FH}, with new geometric and topological  insights 
associated with twists and differential refinements. This leads to  a hierarchy of 
descriptions of these fields and culminating with 
one in twisted differential K-theory, putting on firm ground speculations in the literature and uncovering 
new effects. This crucially uses the series of constructions
by the authors \cite{GS1}\cite{GS3}\cite{GS2}\cite{GS4}\cite{GS5}\cite{GS6}, which to a large
extent were originally motivated by this problem. In some sense then, this is the main application
of the above works. The readers interested in the general mathematical theorems are encouraged to 
consult the above papers, while here we mainly focus on those results that are used in the 
particular physics problem at hand.

\medskip
The RR fields originate as follows.
Introducing fermions into the bosonic string requires considering boundary conditions for these fermions.
Imposing the periodic boundary conditions on the circle, also known as the Ramond boundary conditions,
 leads to the Ramond-Ramond sector, which includes other fields in addition to the spinors \cite{Ra}.
 Among these are the Ramond-Ramond (RR) fields, which are a priori  differential form fields in the 
 10-dimensional spacetime of type II supergravity theories, the latter viewed essentially as the classical 
 limits of type II string theory. 
 
 \medskip
Aside from arising in the spectrum, what is the nature of a Ramond-Ramond field? One can 
actually ask a more basic question: What is a form field in physics? The question might have multiple
answers even when referring to the same field. That is, the mathematical 
description of the field might depend on which aspects of the field one is trying 
to capture. As  in the approach in \cite{Fr}\cite{Fr2}\cite{tcu}, 
 one thematically and schematically has the following picture 
$$
{\tiny
\fbox{RR as Differential form} 
\quad \leadsto \quad 
\fbox{RR in de Rham cohomology} 
\quad \leadsto \quad 
\fbox{RR in integral cohomology} 
\quad \leadsto \quad 
\fbox{RR in generalized cohomology} 
}
$$


As we will see this picture also requires further refinements, including adding periodicity, 
adding a twist as well as adding the data of a connection. On the conceptual side, part of this paper 
 hence also proposes one way of {\it how} to approach answering the above question. Thus, in 
 addition the firm mathematical grounding, we hope to also provide an approach that helps 
 in the conceptual understanding of the problem.

 \medskip
To start, the spectrum of supergravity a priori provides potentials of degrees less than half the dimension 
 of the space. However, in a democratic formulation \cite{To} one would like to have all the RR 
 potentials, while the other half is supplied by a form of Hodge duality. A doubled 
 formalism in which a Hodge dual potential is introduced for each bosonic form field is given in 
 \cite{CJLP}, where the equations of motion can then be formulated as a twisted self-duality 
 condition on the total field strength. Duality-symmetric action for type IIA  is given in \cite{BNS}, 
 with the corresponding duality relations deduced directly from the action.  A generalized form 
 of IIA/IIB supergravity depending on all RR potentials $C_p$, $p=0,1,...,9$, as the effective 
 field theory of Type IIA/IIB superstring theory \cite{BKORvP}. 

\medskip
The RR field strength at the level of supergravity is then an $m$-form $G_m \in \Omega^m(X^{10})$.
The collection of these further occur as even degree forms 
in type IIA and odd degrees in type IIB, up to dimension 10.  This periodicity 
or grading can be taken into account. As explained in \cite{Fr2}, 
one introduces the electromagnetic duals of the supergravity fields
and forms the inhomogeneous RR field strengths
\(
\label{total-RR-form}
G_{\rm form}=\left\{ \begin{tabular}{lll} 
$G_0 + G_2 +  G_4 +  G_6 +  G_8 + G_{10},$&& {Type IIA},
\\
  $G_1 +  G_3 +  G_5 +  G_7 +  G_9$, && {Type IIB},
 \end{tabular}
\right.
\)
where $G_m$ is a differential form of degree $m$. 
Classically, these satisfy appropriate Hodge duality relations.

\medskip
Extracting gauge equivalence classes leads to a description via de Rham cohomology, 
that is 
\(
\label{RR-dR}
\text{RR field}= \{G \in \Omega^\bullet(X^{10})\}/\{G=dC\} \in H_{\rm dR}^\bullet (X^{10}) .
\)
Furthermore, taking into account quantum effects, including Dirac quantization,
leads to a description via integral cohomology. However, as explained in \cite{MW}
this only works for low degrees and under special conditions. Nevertheless taking this
further and requiring the tangent bundle and the gauge bundle to satisfy some congruences, 
as explained in \cite{NS5}, we have that these restricted RR fields are integral cohomology
classes. Taking periodicity and/or twists into account these would be periodic and/or twisted
integral cohomology classes in the sense of \cite{GS4}\cite{GS6}. Differentially refining this
setting means that we are taking these restricted fields and describing them using 
twisted periodic differential integral cohomology, one prominent description of which 
is via twisted periodic Deligne cohomology, constructed in \cite{GS4}\cite{GS6}.

\medskip
Taking into account anomalies and properly accounting for torsion leads to 
RR fields and fluxes being quantized by K-theory \cite{FH}\cite{MW}.
Retaining periodicity via the inverse Bott element $u\in K^{2}({\rm pt})$, 
these can be defined to be homogeneous elements (as in \cite{Fr2})
\(
\label{total-RR-form}
G=\left\{ \begin{tabular}{ll} 
$G_0 + u^{-1}G_2 + u^{-2}G_4 + u^{-3}G_6 + u^{-4}G_8 + u^{-5}G_{10}$, & { Type IIA};
\\
 $u^{-1}G_1 + u^{-2}G_3 + u^{-3}G_5 + u^{-4}G_7 + u^{-5}G_9$, & ~{Type IIB}.
 \end{tabular}
\right.
\)
This element is of degree 0 for IIA and -1 in IIB. For a K-theory class $x$, the resulting quantization 
on a 10-dimensional manifold $X$ takes the form \cite{MW}\cite{FH}
\(
G(x)= \ch (x) \sqrt{\widehat{A}(X)}\;,
\label{RR-field ch}
\)
where $\sqrt{\widehat{A}(X)}$ is the formal square root of the  $\widehat{A}$-genus 
expansion in terms of the Pontrjagin classes, and $\ch: K^*(X) \to H^*(X)$ 
is the Chern character, mapping $K^0$ to even degree cohomology and $K^1$ to odd degree 
cohomology of $X$.

\medskip
Considering a background field or flux changes the system and can be defined at more
than one level.  At the classical level, the fields are just given by differential forms, so a 
background field is a closed 3-form $H$ which leads to modifications of the field equations. 
 More precisely, in the presence of a $B$-field or $H$-flux, the fields satisfy the twisted
 Bianchi identity, which combines what traditionally would be called an
 equation of motion and a Bianchi identity, at the level of forms
 \footnote{In the presence of a B-field we will denote the (rational) RR fields by $F$. These
 are the improved field strengths which are neither closed nor quantized, but are twisted closed.
  }
 $$
 dF_n + H_3\wedge F_{n-2}=0\;. 
 $$
  Using the total field description, 
  this has been
  written succinctly as 
  $$
  d_HF=0\;,
  $$
 where $d_H=d + H_3$ is the twisted differential on the de Rham complex
(see \cite{BCMMS}\cite{MS}\cite{Ev} \cite{tcu}). 
 Hence the fields are then closed under the differential $d_H$  and are classified,
up to equivalence, by the $H$-twisted de Rham cohomology
$$
H^{*}_{\rm dR} (X; H) := \ker(d_H)/ {\rm im}(d_H).
$$

\medskip
When looking at the fields in the presence of a background $H$-flux, one
 needs to extend to twisted setting,  that is, RR fields are 
classified by twisted K-theory. The quantization  condition for the case of a 
twist which is zero in cohomology is  \cite{MoS}
$$
G(x)=e^{B_2} \ch (x) \sqrt{\widehat{A}(X)} \;,
$$
while for a cohomologically nontrivial twist it takes the form \cite{MS}\cite{BMRS}
$$
G(x)=\ch_H \sqrt{\widehat{A}(X)} \in H_H^* (X)  \quad \text{for} \; x \in K_H(X)
$$
where $\ch_H$ is the twisted Chern character for $K_H^*(X)$
 \cite{BCMMS}\cite{MSt}\cite{AS2}\cite{Kar}\cite{HM}. 

\medskip
 Twisted K-theory consistently matches the reduction from M-theory (see \cite{DMW}\cite{MS}) 
and can even be \emph{derived} from M-theory at the rational level (see \cite{FSS16b} for the truncated
case and \cite{vin} for the full case) and beyond the rational level \cite{BSS19}. However, incorporating S-duality 
in type IIB string theory remains a challenge \cite{DMW}\cite{KS2}\cite{BEJMS}\cite{Ev}.

\medskip
Physical considerations generally require one to work with geometric representatives of 
cohomology classes, in the form of differential cohomology (see \cite{Fr}\cite{Fr2}\cite{HS}\cite{Sz}\cite{Urs}\cite{FSS2} for 
motivations and surveys). As our viewpoint involves a hierarchy of descriptions, we start with 
differential integral  cohomology. This is most succinctly described with the 
``differential cohomology diamond diagram" 
\footnote{This diamond (or hexagon) diagram was originally introduced and emphasized by Simons and 
Sullivan in \cite{SSu} and for more generalized theories, a full characterization via this diamond was 
proved in \cite{BNV}. Parts of it appear in the foundational work of Cheeger and Simons \cite{CS}. }
\(
\label{diamond}
\xymatrix{
 &\Omega^{*-1}(M)/{\rm im}(d) \ar[rd]^{a}\ar[rr]^{d} & &
   \Omega^*_{\rm cl}(M)\ar[rd] &  
\\
H^{*-1}_{\rm dR}(M)\ar[ru]\ar[rd] & & 
{\widehat{H}^*(M;\ZZ)}\ar[rd]^{\mathcal{I}} \ar[ru]^{\mathcal{R}}& &  
H^*_{\rm dR}(M) \;,
\\ 
&H^{*-1}(M;\RR/\ZZ) \ar[ru]\ar[rr]^{\beta} 
& & H^*(M;\ZZ) \ar[ru]^{r} &
}
\) 
where $d$ is the de Rham differential, $\mathcal{R}$ is the curvature map, $\mathcal{I}$ is 
the forgetful map, $r$ is  the rationalization, and $\beta$ is the Beckstein associated with the 
exponential coefficient sequence. 
The corresponding description of various facets of the RR fields 
are then captured as 

\(
\label{diamond1}
\xymatrix{
 & 
 \{C\neq dA \} 
\ar[rd]^{a}\ar[rr]^{d} 
& &
  \{G, dG=0\}
  \ar[rd] &  
\\
\{[C]_{\rm dR}\} 
\ar[ru]\ar[rd] & &
 \{[\widehat{G}]\}\ar[rd]^{\mathcal{I}}
  \ar[ru]^{\mathcal{R}}& &  
  \{ [G]_{\rm dR}\} \;,
\\
& \{ [G]_{\rm flat}\} \ar[ru]\ar[rr]^{\beta} & & \{ [G]_{\rm integral}\} \ar[ru]^{r} &
}
\) 
An abelian field represented by a differential K-theory class $F\in \widehat{K}^*(X)$ 
contains the differential form information $\mathcal{R}(F)$. The latter satisfies $d\mathcal{R}(F)=0$ 
in the absence of D-brane sources, which is what  we are assuming here. 
The de Rham class represented by $\mathcal{R}(F)$ is  quantized to lie 
in an integral lattice given by the image of the Chern character, as 
 $F$ also contains the integral
(and possibly torsion) information of the class $\mathcal{I}(F) \in K^*(X)$. 
Note that $\mathcal{I}(F)$ and $\mathcal{R}(F)$ together do not determine $F$ entirely, 
as one needs to supply the extra information,
corresponding to a potential with corresponding gauge transformations.

\medskip
Differential K-theory as the home for RR fields without H-field has been 
advocated in \cite{FH}\cite{Fr} \cite{Fr2}. 
The need for twisted differential K-theory for description of the fields in string with 
an H-field  has been highlighted in \cite{Fr}\cite{FMS}\cite{BeM}\cite{KV} for general 
classical backgrounds and in \cite{DFM} for  orientifolds.  
Characterizations of various aspects of twisted differential  K-theory are 
given in \cite{CMW}\cite{KV}\cite{BN}, culminating most concretely for our purposes in 
\cite{GS5}. In fact, one of the original 
motivations for constructing the latter as the last in a series of papers was to generally 
provide a proper receptacle for the RR fields in the presence of  
twisting NS fields. The theory sits in the following diagram 
$$
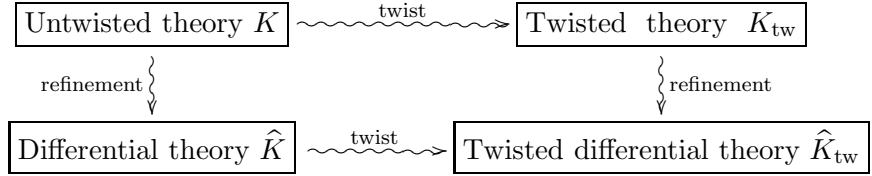
\xymatrix{
\fbox{\text Untwisted  theory $K$}  \ar@{~>}[rr]^-{\rm twist} \ar@{~>}[d]_-{\rm refinement} &&
\fbox{Twisted~ theory~ $K_{\rm tw}$} \ar@{~>}[d]^-{\rm refinement} 
\\
\fbox{\text Differential theory $\widehat{K}$} \ar@{~>}[rr]^-{\rm twist} && 
\fbox{\text Twisted differential theory $\widehat{K}_{\rm tw}$}
}
$$

\medskip
What is needed to fully describe RR fields explicitly?
The general approach emphasized in \cite{KS1} is to view physical conditions 
as obstructions to orientation, or as differentials in the Atiyah-Hirzebruch spectral sequence (AHSS), 
extending even beyond K-theory. In that direction, our work establishes in the series 
\cite{GS1}\cite{GS2}\cite{GS3}\cite{GS4}\cite{GS5}\cite{GS6} 
that the AHSS  can be extended to the differential refinements, 
that is, we can refine the AHSS for an untwisted or twisted topological theory $E$, such as K-theory,
 by appropriately adjoining geometric data to it.  With our explicit descriptions of the differentials in 
the AHSS for twisted differential K-theory, we are able to make such a general description 
manifest and precise, supplying the missing ingredients that lead 
to a more complete picture than was previously possible.  

\medskip
In the above works we established the differential refinement of the following ingredients which
enter into the picture. Note that the differentials in the untwisted AHSS are primary  operations, 
and the ones in the twisted theory are secondary operations \cite{AS2}.
\begin{enumerate}
\vspace{-1mm}
\item Differential refinement of {\color{blue} primary} cohomology operations: 
{\color{blue} Steenrod squares} $Sq$. 
\vspace{-2mm} 
\item Differential refinement of {\color{green} secondary} cohomology operations:
 {\color{green} Massey products} 
$\langle \cdot , \cdot , \cdots  \rangle_{{}_{\rm Massey}}$. 

\vspace{-2mm} 
\item Differential refinement of  the AHSS with a concrete identification of the differentials, such that 
we have the following diagram for the differentials

\vspace{-4mm}
$$
\xymatrix{
{\color{blue} d} \ar@{~>}[rr]^-{\rm twist} \ar@{~>}[d]_-{\rm refinement} 
&&
d_{\rm twist}= {\color{blue} d} + {\color{green}  \rm secondary\ operation}
 \ar@{~>}[d]^-{\rm refinement} 
 \\
d_{\rm refined}= \widehat{\color{blue} d}
\ar@{~>}[rr]^-{\rm twist} &&
\widehat{d}_{\rm twist}= \widehat{\color{blue} d} + \widehat{\color{green} \rm secondary\ operation}
}
$$
Note that, as explained in our work above, differentially refining twisted K-theory is equivalent to
twisting differentially refined K-theory, i.e., `$[\text{twisted}, \text{differential}]=0$'. 

\end{enumerate}

\medskip
The AHSS for twisted differential  K-theory \cite{GS5} will be denoted by $\widehat{\rm AHSS}_{\hat \tau}$, 
where $\hat{\tau}$ is a representative of a differential cohomology class, i.e. a higher bundle with 
connection. When this twisting class is zero in differential cohomology, 
we recover the AHSS for differential K-theory  \cite{GS3}, which
we denote $\widehat{\rm AHSS}$. On the other hand, if we forget the differential 
refinement and reduce the theory to its underlying topological content then 
we recover the AHSS for twisted K-theory constructed by Rosenberg and Atiyah-Segal
\cite{Ro}\cite{AS2},  which we denote ${\rm AHSS}_{\tau}$. 
When we take both a trivial twist and no differential refinement, then we restrict to the original case 
considered by Atiyah and Hirzebruch \cite{AH2}. As explained in \cite{GS5}, we overall have a 
correspondence diagram of transformations of the corresponding spectral sequences 
$$
\xymatrix{
 \widehat{\rm AHSS}_{\hat \tau} \ar@{~>}[d]_{\hat \tau=0} \ar[rrr]^{\mathcal{I}_{\tau}} 
 &&& {\rm AHSS}_{\tau} \ar@{~>}[d]^{\tau=0} \\
 \widehat{\rm AHSS} \ar[rrr]^{\mathcal{I}}&&&
 {\rm AHSS}
}
$$
where $\tau$ is the twist and $\mathcal{I}$ is the reduction to the topological part.

\medskip
 From a physics perspective, it is important to determine when  a cohomology class $x \in H^i(M; \ZZ)$ 
lifts to a class $\alpha (x)$  in K-theory $K(M)$ (see \cite{DMW}). The obstruction  is given 
by the differential $d=Sq^3$ in the AHSS, i.e., it is a necessary condition that $Sq^3 x=0$. Likewise,  
for twisted K-theory, the obstruction is $Sq^3x + H\cup x=0$, which is again the differential in the 
$\op{AHSS}_H$ (see \cite{DMW}\cite{ES}\cite{BEJMS}).  We would like to extend this 
to the twisted differential case.

\medskip
A similar argument holds from the  homological point of view of branes  ending on other 
branes \cite{MMS}\cite{BEJMS}. 
Anomalies associated with D-branes in the presence of a B-field have been considered
in \cite{FW}. This involves three factors, the holonomy of the B-field over the 2-dimensional 
string worldsheet,  the holonomy of the Chan-Paton bundle
along the boundary of the string, and the Pafaffian associated with the path integral of the 
spinors. None of these factors are globally well-defined, leading to a description of the partition 
function as a section of a tensor product of three line bundles. The nontriviality of the resulting 
line bundle is the Freed-Witten anomaly and the necessary condition for the anomaly to vanish is the 
Freed-Witten anomaly cancellation condition $W_3 + H_3=0$. This has been generalized to 
the case when the two classes differ by a torsion class \cite{Kap}, studied from the point of  
view of gerbes in \cite{CJM}\cite{BFS}, interpreted as a pushforward in twisted K-theory in \cite{CW}\cite{ABG},
and described via higher geometric quantization and smooth stacks in \cite{FSS2}. 
What we would like to establish is the following:
\begin{enumerate}
\vspace{-2mm}
\item A differential analogue of the Freed-Witten 
condition \cite{FW}, i.e.,
$$
\widehat{W}_3 + \widehat{H}_3=0\;.
$$

\vspace{-4mm}
\item An interpretation as a differential in twisted differential K-theory $\widehat{\rm AHSS}_{\hat \tau}$. 
\end{enumerate}
Sufficiency in the presence of branes, involving Steenrod power operations 
at odd primes in the context of Steenrod's problem on realization of  homology classes as submanifolds
is discussed in \cite{ES}.

\medskip
We are also interested in finding a twisted differential version of the quantization condition \eqref{RR-field ch} on 
the RR fields. Earlier attempts include the following. Using the language of differential characters, in 
\cite{BeM2} a version of the 
twisted differential Chern character was proposed with 
$
\hat{G}(\hat{x}) = \sqrt{\widehat{A}(X,\nabla_{g})} {\ch}_{\hat{B}}(x)
$, where $\nabla_g$ is the metric connection and $\hat{B}$ is a flat character. 
It was also speculated in \cite{KM} that the quantization condition 
for the RR fields in differential K-theory takes the form (at the level 
of differential forms) would be
$
\hat{G}(\hat{x}) = \sqrt{\widehat{A}(X)} {\ch}(\hat{x}) 
$,
for $\hat{x}\in \widehat{K}(X)$ while in the twisted case, the 
only effect of the B-field was to modify the connection entering in the 
form representative of the $A$-genus (which is argued why it is not modified).
We will define the proper expression  and  make good mathematical sense of the quantity
\(\label{rrfieltwdf}
\widehat{G}(\hat{x})=\widehat{\ch}_{\hat{h}}(\hat{x})\cup_{\rm DB} \sqrt{\widehat{\mathbb{A}}(X,\nabla_{g})} \;,
\)
as a differential cohomology class, where $\cup_{\rm DB}$ is the Deligne-Beilinson cup product, which is in a sense
an extension of the cup product to differential cohomology  
(see \cite{cup}\cite{FSS2}). This involves, for every $U(1)$-gerbe with 
connection $\hat{h}:X\to \BB^2U(1)_{\nabla}$, a generally defined twisted differential Chern character
\(\label{dfchc1}
\widehat{\ch}_{\hat{h}}:\widehat{K}^*_{\hat{h}}(X)\longrightarrow \widehat{H}^*_{\hat{h}}(X;\QQ[u,u^{-1}]) \;.
\)
The source of the of the character is twisted differential K-theory and the target is a differential refinement of twisted
 \emph{periodic rational} cohomology, i.e., considering rational cohomology rolled up into 
 even and odd degrees. For general twisting classes, making sense of the map \eqref{dfchc1} is highly nontrivial.
 In the special case where the twist is torsion, one can make good sense of the twisted differential Chern character 
 via concrete models (see  \cite{CMW}\cite{Pa}). In the general case this has only been put on firm ground recently (see \cite{BN}\cite{GS5}). 
The quantity \eqref{rrfieltwdf} also involves a differential refinement of the $\hat{A}$-genus, which we have 
addressed in detail in \cite{GS7}. 

\medskip
Note that, as indicated right after \eqref{RR-dR} above, in specialized settings (e.g. those orientifolds where 
K-theoretic effects might not be  seen) one might consider, for instance, integral 
cohomology. This then  leads to twisted integral cohomology with twist given 
as a mod 2 degree one class, as constructed in \cite{GS4}.
Here one again tries to lift to twisted de Rham cohomology, obtained by 
 those RR fields that are $d_H$-closed modulo $d_H$-exact, twisting 
 \eqref{RR-dR}.

 \begin{center}
 \begin{tabular}{|c||c|}
 \hline
{\bf  Mathematical description} & {\bf Physical setting}
 \\
 \hline
 \hline
 1-twisted integral cohomology & Orientifold fields 
 \\
 \hline
 1-twisted Deligne cohomology & Differential orientifold fields 
 \\
 \hline
 \end{tabular} 
 \end{center}
  Alternatively, we can also consider a higher-degree twist (including three) for a periodic version of 
Deligne cohomology \cite{GS6}, which can be viewed as a twisted extension of approaches via 
differential cohomology or differential characters (see e.g. \cite{BeM2}\cite{Monn}). 
While we do not pursue this explicitly and in detail here,
we find it useful to point them out as sort of  intermediate cases between 
twisted de Rham cohomology and twisted K-theory, in the sense of the following tables. 

\vspace{-2mm}
\begin{center}
\begin{tabular}{|c|c|c|}
\hline
{\bf Field as element of} & {\bf Twist} & {\bf Field + twist} \\
\hline
\hline
de Rham cohomology & closed 3-form/de Rham 3-class & Twisted periodic de Rham cohomology\\
\hline
Deligne cohomology & gerbe & Twisted Deligne cohomology \\
\hline
K-theory & integral 3-class & Twisted K-theory \\ 
\hline
Differential K-theory & gerbe & Twisted differential K-theory \\ 
\hline
\end{tabular}
\end{center}

  \begin{center}
 \begin{tabular}{|c||c|}
 \hline
{\bf  Mathematical description} & {\bf Physical setting}
 \\
 \hline
 \hline
 3-twisted periodic integral cohomology &   Integral fields 
 \\
 \hline
 3-twisted periodic Deligne cohomology & Differential fields 
 \\
 \hline
  Twisted differential K-theory & General RR fields 
 \\
 \hline
 \end{tabular} 
 \end{center}
 One could also 
consider higher theories beyond K-theory \cite{KS1}\cite{KS2}\cite{KS3}\cite{tcu}\cite{SW}\cite{LSW}.
However, we will leave this for a separate discussion and focus here on K-theory.

\medskip
The paper is organized as follows. 
We describe the general setting of twisted differential K-theory as the receptacle for the RR fields
in Sec. \ref{Sec-quan}, recalling constructions and results from earlier work. We start with 
differential K-theory in Sec. \ref{Sec-twisteddiffk} and then twisted K-theory in Sec. \ref{Sec-toptwisted},
combining the two  into twisted differential K-theory in Sec. \ref{sec-diff-twk}, with the main 
 highlight being the twisted differential Chern character and the refinement of the $\widehat{A}$-genus. 
 This then leads to a justification of why \eqref{rrfieltwdf} is the right definition.
In Sec. \ref{Sec-lift}  we study the lifting of RR differential forms to twisted differential K-theory, with the main
tool being the twisted differential AHSS. This involves determining explicitly in Sec. \ref{Sec-liftK} 
the torsion differentials and identifying obstructions associated to both flat classes and curvature 
forms. The detailed analysis leads to shifted quantization conditions on the fields with the highlight being
an explicit and detailed algorithm for characterizing and detecting  RR fields. 

\medskip
Moving to the twisted case in Sec. \ref{Sec-liftKt}, we describe the dynamics of the 
twisted RR fields via Massey products, also finding the higher potentials for the Massey products
themselves. We then identify the higher differentials in the $\widehat{\rm AHSS}_\tau$ 
via the the differentially refined Massey products from our earlier work, and determine conditions for lifting flat classes to
 twisted differential K-theory. Then we consider the anomalies in Sec. \ref{anomalies},
where we provide our refinement of the Freed-Witten anomaly. Finally, in the last 
section, Sec. \ref{Sec-comp}, we illustrate the description of RR fields in nontrivial backgrounds
by calculating the twisted differential K-theory for prominent examples of importance to 
type II string theory, namely spheres in Sec. \ref{Sec-sph}, tori in Sec. \ref{Sec-tori},
and Calabi-Yau threefolds $CY_3$ and to some extent compact 6-dimensional manifolds  
in Sec. \ref{Sec-CY}. The latter generalizes and extends 
results of Doran and Morgan \cite{DM} who computed the topological K-theory of 
such manifolds.

\section{RR fields as twisted differential K-theory classes}
\label{Sec-quan}

\subsection{Differential K-theory}
\label{Sec-twisteddiffk}

In this section, we review the Hopkins-Singer type differential $K$-theory \cite{HS}, presented as a sheaf of 
spectra \cite{BNV}\cite{Urs}.  The material in this section is well-known to the experts; nevertheless, because 
the machinery is highly technical, we  have decided  to review the construction briefly here. For the reader 
who is not interested in these technicalities,  this section can be safely skipped.

\medskip
We consider topological (smooth) spaces as modeled using (smooth) infinity groupoids, i.e., as objects in $\infty\mathscr{G}{\rm pd}$.
Let $C\mathscr{M}{\rm on}(\infty\mathscr{G}{\rm pd})$ denote the sub $\infty$-category of commutative monoids in 
$\infty\mathscr{G}{\rm pd}$, and let $C\mathscr{G}{\rm rp}(\infty\mathscr{G}{\rm pd})$ be the 
subcategory of $\infty$-abelian  groups (i.e. connected spectra). The inclusion 
$i:C\mathscr{G}{\rm rp}(\infty\mathscr{G}{\rm pd})\into C\mathscr{M}{\rm on}(\infty\mathscr{G}{\rm pd})$
 admits a left adjoint $\mathscr{K}$ which can be thought of as taking the group completion. The functor 
 $\mathscr{K}$ prolongs to a functor between presheaves of $\infty$-monoids and $\infty$-abelian groups. 

\begin{definition}[Smooth K-theory spectrum]
Let $L:\mathscr{P}\mathscr{S}{\rm h}_{\infty}(\mathscr{M}{\rm an};\mathscr{S}{\rm p})\to \mathscr{S}{\rm h}_{\infty}(\mathscr{M}{\rm an};\mathscr{S}{\rm p})$ denote the stackification functor (left adjoint to the inclusion $i$). We define the smooth $\op{KU}$-spectrum with 
connections as the connected sheaf of spectra defined by 
$$
{\bf k}{\rm U}:=L\circ \mathscr{K}\Big(\coprod_{n\in \NN}\BB {\rm U}(n)_{\nabla}\Big).
$$
\end{definition}

\begin{remark}[Vector bundles with connections]
Note that it is immediate from the definition (see \cite{GS7} for the real case) that we have a natural isomorphism
$$
{\bf k}{\rm U}_{\nabla}(M)\cong {\rm Gr}\big({\rm Vect}_{\nabla}^g(M)\big),
$$ 
where ${\rm Vect}_{\nabla}^g(M)$ is the category of vector complex vector bundles with Hermitian metric connections (with isomorphisms 
between them) and ${\rm Gr}$ denotes the Grothendieck group completion.
\end{remark}

The construction of the Hopkins-Singer refinement of the $K$-theory spectrum proceeds by applying the cohesive $\infty$-adjoints $(\delta^{\dagger}\vdash \Gamma \vdash \delta \vdash \Pi)$ as introduced in \cite{Urs} 
to the sheaf of spectra ${\bf k}{\rm U}_{\nabla}$. The topological realization 
\footnote{Thes composite functor takes what is traditionally called the \emph{geometric realization} of the sheaf of spectra and then embeds it as a  constant sheaf of spectra. However, the term geometric here is misleading, as the result is a topological space. Hence
 we have opted to call this operation the \emph{topological realization}.}
$\delta\Pi$ induces a morphism of sheaves of spectra (see \cite{BNV})
\(\label{cycmp}
{\rm cyc}:=\delta\Pi:\mathbf{k}{\rm U}_{\nabla}\simeq \op{kU}\longrightarrow {\rm KU} .
\)
We will need the following ingredients:

\begin{enumerate}
\vspace{-3mm}
\item[{\bf (i)}] The Eilenberg-MacLane functor 
$
H:\mathscr{C}{\rm h}\longrightarrow \mathscr{S}{\rm p}
$,
which sends an unbounded chain complex to a corresponding spectrum. 
\vspace{-3mm}
\item[{\bf (ii)}] Let $\Omega^*(-;\pi_*(\op{K}))$ be the complex of forms with coefficients in $\pi_*(\op{K})$. 
Explicitly, by rationalizing the  coefficients of K,  this complex is 2-periodic and looks as follows
$$
\Omega^*(-;\pi_*(\op{K}))=\big(\vcenter{\xymatrix@=1.5em{
\hdots \ar[r] &\bigoplus_n \Omega^{2n}\ar[r] & \bigoplus_n \Omega^{2n+1}\ar[r] & \bigoplus_n \Omega^{2n}\ar[r] & \bigoplus_n \Omega^{2n+1}\ar[r] & \hdots 
}}\big).
$$

\vspace{-6mm}
\item[{\bf (iii)}] We can truncate the complex $\Omega^*(-;\pi_*(\op{K}))$ at degree zero, removing 
all forms in negative degrees. We denote this truncated complex by 
$$
\tau_{\leq 0}\Omega^*(-;\pi_*(\op{K}))=\big(\vcenter{\xymatrix{
\hdots \ar[r] & 0 \ar[r] &  0 \ar[r] & \bigoplus_n \Omega^{2n}\ar[r] 
& \bigoplus_n \Omega^{2n+1}\ar[r] & \bigoplus_n \Omega^{2n} \ar[r] & \hdots 
}}\big),
$$
where the first nonzero component appears in degree zero. 
\vspace{-3mm}
\item[{\bf (iv)}] The Chern character form gives a morphism of smooth stacks (preserving the monoidal structure)
$$
\ch:\coprod_{n\in \mathbb{N}}\mathbf{B}{\op{U}}(n)_{\nabla}
\xymatrix{\ar[r]&} \Omega^0(-;\pi_*(\op{K})).
$$
\end{enumerate} 
Since $\Omega^0(-;\pi_*(\op{K}))$ is already a sheaf of abelian groups, we have 
$\mathscr{K}(\Omega^0(-;\pi_*(\op{K})))=H(\Omega^0(-;\pi_*(\op{K})))$. Postcomposing with the canonical map 
$
i^*: H(\Omega^0(-;\pi_*(\op{K}))) \to
H(\tau_{\leq 0}\Omega^*(-;\pi_*(\op{K})))
$,
induced by the inclusion $i:\Omega^0(-;\pi_*(\op{K}))\into \tau_{\leq 0}\Omega^*(-;\pi_*(\op{K}))$, 
we get an induced map on completions
\(\label{pchrefsp}
\ch:\mathbf{k}{\op{U}}_{\nabla}:=
\mathscr{K}\Big(\coprod_{n\in \mathbb{N}}\mathbf{B}{\op{U}}(n)_{\nabla}\Big)\xymatrix{\ar[r]&}
H(\tau_{\leq 0}\Omega^*(-;\pi_*(\op{K}))).
\)
Geometrically (more properly, topologically) realizing this map and using \eqref{cycmp} gives rise to a map
$
\widetilde{{\rm ch}}:\op{K}\xymatrix{\ar[r]&} H(\mathbb{R}[u,u^{-1}]) ,
$,
where $u$ is the Bott periodicity element of degree $\vert u\vert=2$. 

\begin{definition} [Hopkins-Singer differential K-theory]
\item{\bf (i)} The differential K-theory spectrum is defined via the pullback in sheaves of spectra
$$
\xymatrix@R=1.7em{
{\rm diff}\big({\rm K}, \widetilde{\ch}, \pi_*(\op{K})\big)
\ar[r]\ar[d] & H\big(\tau_{\leq 0}\Omega^*(-;\pi_*(\op{K}))\big)
\ar[d]
\\
\op{K}\ar[r]^-{\widetilde{\ch}} &  H(\pi_*(\op{K})\otimes \RR)
}\!.
$$
This pullback depends on the map $\widetilde{\ch}$ and the graded ring $\pi_*({\rm K})$. 
We fix this data once and for all and denote the sheaf of spectra simply as
$$
\widehat{\rm K}:= {\rm diff}\big({\rm K}, \widetilde{\ch} ,\pi_*(\op{K})\big).
$$
{\bf (ii)} The differential K-spectrum refining higher degree K-groups is given by the pullback 
$$
\xymatrix@R=1.7em{
{\rm diff}\big(\Sigma^n{\rm K},\Sigma^n(\widetilde{\ch}), \pi_*(\op{K})[n]\big)
\ar[r]\ar[d] & H\big(\tau_{\leq 0}\Omega^*(-;\pi_*(\op{K})[n])\big)\ar[d]
\\
\Sigma^n\op{K}\ar[r]^-{\Sigma^n(\widetilde{\ch})} &  H(\pi_*(\op{K})[n]\otimes \RR)
}\!.
$$
where $\Sigma^n$ denotes the $n$-fold suspension and $\pi_*(\op{K})[n]$ denotes the shift of the complex 
$\pi_*(\op{K})$ up $n$-units. Again we fix this data once and for all and define 
\footnote{These sheaves of spectra are not to be confused with the notation for homology, which we do not consider in 
this paper.} 
$$
\widehat{\rm K}_n:= {\rm diff}\big(\Sigma^n{\rm K},\Sigma^n(\widetilde{\ch}), \pi_*(\op{K})[n]\big).
$$
\item {\bf (iii)} Differential K-cohomology of a manifold $M$ is defined as 
$$
\widehat{\rm K}^n(M):=\pi_0\map(M, \widehat{\rm K}_n).
$$
\end{definition}

One has the following properties, as for any differential cohomology theory.   

\begin{remark}
[Basic properties of $\widehat{\rm K}$]
\item [{\bf (i)}] {\rm (Diamond)}
From \cite[Lemma 6.8]{BNV}, we see that the differential cohomology hexagon diagram takes the following form
\(
\label{kodfdiam}
\xymatrix @C=-34pt @!C{
&\Omega^{*-1}(M;\pi_*(\op{K}))/{\rm im}(d) \ar[rd]^{a}\ar[rr]^{d} & &
\Omega_{\rm cl}^*(M;\pi_*(\op{K}))\ar[rd] &  
\\
H^{*-1}(M;\pi_*(\op{K})\otimes \RR)\ar[ru]\ar[rd] & &\widehat{\rm K}(M)\ar[rd]^{\mathcal I} \ar[ru]^{\mathcal R}& &  H^*(M;\pi_*(\op{K})\otimes \RR) \;,
\\
&\op{K}^{*-1}(M;U(1))\ar[ru]^-j\ar[rr]^{\beta} & & \op{K}^*(M) \ar[ru] &
}
\) 
which related differential K-theory to the underlying topological theory and differential form representatives for the rationalization. 
\item [{\bf (ii)}] {\rm (Coefficients)}
Both diagonals in the diagram are exact and the bottom sequence is exact -- induced from the cofiber/fiber sequence
$$
\op{K}\simeq \op{K}\wedge \mathbb{S}\longrightarrow \op{K}\wedge \mathbb{S}\RR\longrightarrow 
\op{K}\wedge \mathbb{S}U(1) ,
$$
where $\mathbb{S}\RR$ and $\mathbb{S}U(1)$ are Moore spectra for $\RR$ and $U(1)$, respectively. 
These correspond to the cohomology theories with coefficients, namely 
${\rm K}^*(-)$, ${\rm K}^*(-; \R)$, and ${\rm K}^*(-; U(1))$, respectively. 

\item [{\bf (iii)}] {\rm (Mayer-Vietoris)}
Again applying the general construction of \cite{BNV} to our case, if $M$ a smooth manifold  and $\{U,V\}$ an open cover, we also have a Mayer-Vietoris sequence
$$
\xymatrix@R=1.6em{
\cdots \ar[r] & \op{K}^{n-2}(U\cap V;U(1)) \ar[r] & \widehat{\rm K}^n(M) \ar[r]&\widehat{\rm K}^n(U)\oplus \widehat{\rm K}^n(V)
\ar@{->} `r/8pt[d] `/12pt[l] `^d[ll]+<-10ex> `^r/8pt[dll] [dll] 
\\
&  \widehat{\rm K}^n(U\cap V)\ar[r] & \op{K}^{n+1}(M)\ar[r] & \cdots .
}
$$
\end{remark}

\subsection{The topological twisted K-theory} 
\label{Sec-toptwisted}

For any commutative ring spectrum $\mathscr{R}$, there is a well-defined topological space of invertible module 
spectra ${\rm Pic}(\mathscr{R})$. Heuristically, the elements are invertible module spectra over $\mathscr{R}$ and 
the paths are equivalences of module spectra, etc. More precisely, this is defined as the maximal $\infty$-groupoid 
inside the full sub $\infty$-category of $\mathscr{R}\mathscr{M}{\rm od}$ on invertible objects. This space is related 
to the space of twists $BGL_1(\mathscr{R})$ considered in \cite{May} by 
$$
\Omega {\rm Pic}(\mathscr{R})\simeq GL_1(\mathscr{R}).
$$
Hence the connected component of the identity is equivalent to $BGL_1(\mathscr{R})$. This perspective on the 
twists for a cohomology theory is essentially the $\infty$-categorical treatment taken in \cite{ABGHR}. In general, 
a twist of a ring spectrum $\mathscr{R}$, over a space $X$, is simply a map $h:X\to {\rm Pic}(\mathscr{R})$. 
This twist can be thought of in two dual ways: it defines a higher-categorical local system, given by associating to each 
point $x\in X$ an invertible module spectrum $\mathscr{R}_x$, and to each path an equivalence between two 
such spectra, etc.; alternatively, the $\infty$-Grothendieck construction \cite[Section 3.2]{Lur} (see \cite{GS7} for the construction in the stable case) allows us to construct a canonical bundle of spectra $\xi \to {\rm Pic}(\mathscr{R})$, the fiber 
over an element being given by the invertible module represented by that element. The pullback bundle by a twist
\(
\xymatrix@R=1.5em{
\mathscr{R}_h\ar[rr]\ar[d] && \xi\ar[d]
\\
X\ar[rr]^-{h} && {\rm Pic}({\mathscr{R}})
}
\label{eq-xi}
\)
has fibers canonically identified with the spectrum $\mathscr{R}_x$, which $h$ associated to the point $x$. 
This generalizes the classical duality between local systems and covering spaces. 

\medskip
The $\infty$-categorical machinery described above is quite powerful and can be generalized to differential cohomology 
theories in a fairly natural way \cite{BN}\cite{GS7}. We first illustrate how to utilize this machinery in order to construct 
the twisted Chern character and then generalize to the differential setting. The relevant spectra we will need in order to discuss 
the twisted Chern character are the $K$-theory spectrum and a periodic spectrum generalizing rational cohomology. The latter 
spectrum is constructed as follows. Let $\QQ[u,u^{-1}]$ be the graded algebra with $u$ in degree 2. For every such algebra there 
is an associated Eilenberg-MacLane spectrum $H\QQ[u,u^{-1}]$. This spectrum represents cohomology with coefficients in 
$\QQ[u,u^{-1}]$ and we have an isomorphism of groups
$$
H^0(X;\QQ[u,u^{-1}])\cong \bigoplus_{k\geq 0}H^{2k}(X;\QQ), \qquad 
 H^1(X;\QQ[u,u^{-1}])\cong \bigoplus_{k\geq 0}H^{2k+1}(X;\QQ).
$$
Algebraically, the elements of $H^*(X;\QQ[u,u^{-1}])$ are polynomials in $u$ with coefficients in $H^*(X;\QQ)$, graded 
according to the  parity of the coefficients. Since $\QQ[u,u^{-1}]$ admits the structure of a graded ring, $H\QQ[u,u^{-1}]$ 
admits the structure of a ring spectrum. The K-theory spectrum $\op{K}$ also admits a commutative ring structure 
\cite{May} (see also \cite{Schw}) and the usual Chern character map 
$$
\ch:\op{K}\longrightarrow H\QQ[u,u^{-1}]
$$
defines a map of ring spectra, which rationalizes to an equivalence. The following proposition shows how  to 
produce the twisted Chern character using the machinery of \cite{ABGHR}.

\begin{proposition}[Twisted Chern character]
Let $X$ be a CW-complex and let $h:X\to K(\ZZ,3)$ be a twist for K-theory. Let $K_h\to X$ and $H\QQ[u,u^{-1}]_h\to X$ 
be the bundles of spectra representing $h$-twisted K-theory and rational cohomology (twisted by the post-composition of $h$ 
with the canonical map $K(\ZZ,3)\to K(\QQ,3)$). There is a morphism of bundles of spectra
$$
\ch_h:K_h\longrightarrow H\QQ[u,u^{-1}]_h,
$$
inducing a twisted Chern character map $\ch_h:K_h(X)\to H_h(X;\QQ)$, which reduces to the untwisted 
Chern character when $h$ is trivial.
\end{proposition}
\theproof
We will construct this map universally. The morphism of ring spectra $\ch:\op{K}\to H\QQ[u,u^{-1}]$ induces an 
$\infty$-functor  on the $\infty$-category of modules via 
$\overline{\ch}:\mathcal{L}\mapsto \mathcal{L}\wedge_{\op{K}}H\QQ[u,u^{-1}]$, where
$\mathcal{L}$ is an invertible module spectrum over $\op{K}$.
It is easy to show that this functor preserves the property of invertibility and sends equivalences to equivalences. 
Thus, we have an induced map 
$$
\overline{\ch}:{\rm Pic}(\op{K})\longrightarrow {\rm Pic}(H\QQ[u,u^{-1}]).
$$
But such a map canonically induces a morphism of the corresponding universal bundles of spectra
(see \eqref{eq-xi}) 
$$
\xymatrix{
\xi\ar[rr]^-{\ch}\ar[d] && \xi^{\prime}\ar[d]
\\
{\rm Pic}(\op{K})\ar[rr]^-{\ch} && {\rm Pic}(H\QQ[u,u^{-1}])
,
}
$$
i.e., $\ch$ restricts fiberwise to the map $\overline{\ch}$ described above. Given a twist 
$h:X\to {\rm Pic}(\op{K})$, the 
universal property of the pullback induces a map on corresponding pullback bundles
$$
\ch_h:\op{K}_{h}\longrightarrow H\QQ[u,u^{-1}]_{h^*\ch}.
$$
In addition, it immediately follows from the construction that this reduces to the usual Chern character 
(up to equivalence)  for a nullhomotopic twist $h:X\to \ast\to {\rm Pic}(\op{K})$, where the second map 
in the composite picks out $\op{K}$, which is trivially a module over itself. Thus it only remains to show 
that for a twist of the form  $h:X\to K(\ZZ,3)\into BGL_1(\op{K})\into {\rm Pic}(\op{K})$, the 
induced twist $h^*\ch$ factors through the  rationalization $K(\ZZ,3)\to K(\QQ,3)\into BGL_1(H\QQ[u,u^{-1}])$. 
This can be shown, for example, using Snaith's theorem,  which identifies $K(\ZZ,2)$ in $\op{K}$ with the 
localization of $\Sigma^{\infty}_+K(\ZZ,2)$ at the Bott element \cite{Sn}.  Under this presentation of $\op{K}$, 
the Chern character can be identified with the localization of the rationalization  map 
$\Sigma^{\infty}_+K(\ZZ,2)\to \Sigma^{\infty}_+K(\QQ,2)$.
 \endofproof
 
 The Chern character for twisted K-theory has also been considered via explicit models 
 in various places, including \cite{BCMMS}\cite{MSt}\cite{HM} (see \cite{GS5} for an extensive list of references). 

\subsection{ The differential twisted K-theory}
\label{sec-diff-twk}

We now enhance the previous discussion to twisted differential $K$-theory. Following \cite{BN}, and discussed more 
at-length in \cite{GS7}, we observe that much of the machinery of \cite{ABGHR} can be extended to the setting of differential 
ring spectra. For a differential refinement $\widehat{\mathscr{R}}=(\mathscr{R}, \ch, A)$ of a ring spectrum $\mathscr{R}$, 
equipped with an equivalence $\ch:\mathscr{R}\wedge H\RR\simeq HA$, the smooth stack of twists 
$\widehat{\rm Tw}(\widehat{\mathscr{R}})$ is defined as the (homotopy) pullback 
\(\label{diftwrgsp}
\xymatrix{
\widehat{\rm Tw}(\widehat{\mathscr{R}})\ar[rr] \ar[d]&& {\rm Pic}^{\rm fl}(\Omega^*(-;A))\ar[d]
\\
\underline{{\rm Pic}}(\mathscr{R}) \ar[rr] && {\rm Pic}(H\Omega^*(-;A))
}
\)
where ${\rm Pic}^{\rm fl}(\Omega^*(-;A))$ represents the twists of the de Rham complex with coefficients in $A$ (i.e. invertible 
K-flat modules over $\Omega^*(-;A)$), $\underline{{\rm Pic}}(\mathscr{R})$ is the locally constant stacks on the corresponding 
Picard infinity groupoid and ${\rm Pic}(H\Omega^*(-;A))$ is the stack of locally constant invertible modules over the sheaf of 
spectra $H\Omega^*(-;A)$. These ingredients are described in detail in \cite{BN} and \cite{GS7}. 

\medskip
The case of K-theory is particularly illuminating here. In \cite{GS7}, we showed that differential K-theory can indeed be twisted 
by gerbes with connection. In fact, there is a canonical map 
\(\label{twgerbwcon}
i:\BB^2U(1)_{\nabla}\longrightarrow\widehat{\rm Tw}(\widehat{\op{K}})\;,
\)
where $ \widehat{\rm Tw}(\widehat{\op{K}})$ is the smooth stack of twists for differential K-theory. This map is roughly 
defined as follows (see \cite{GS7} for details). We recall that the stack of gerbes with connections 
$\BB^2U(1)_{\nabla}$ fits into a homotopy pullback (see for example \cite[Section 4.4.15]{Urs}) 
\(\label{diftwkth}
\xymatrix@C=4em@R=1.5em{
\BB^2U(1)_{\nabla}\ar[r] \ar[d]& \Omega^3_{\rm cl}\ar[d]
\\
\BB^3\ZZ \ar[r] & \BB^3\RR
\;.}
\)
Comparing this with the pullback \eqref{diftwrgsp}, we see that we can get an induced map by via the universal property, provided 
we produce maps and homotopies between the corresponding span diagrams. On each manifold $M$, the underlying topological 
twist is defined by $h:M\to \BB^3\ZZ\simeq \underline{K(\ZZ,3)}$, where  $\underline{K(\ZZ,3)}$ is the locally constant stack 
associated to the  ${K(\ZZ,3)}$. This twist   is  regarded as the twist for topological K-theory, the 3-form 
curvature $H\in \Omega_{\rm cl}^3(M)$ is mapped to the twisted de Rham complex $(\Omega^*[u,u^{-1}],d_H)$ on $M$, and the 
homotopy filling the diagram is sent to a twisted de Rham equivalence
$$
d:H((\Omega^*[u,u^{-1}],d_H))\xrightarrow{\simeq} H\RR[u,u^{-1}]_h\;.
$$
The universal property of the homotopy pullback then induces a map \eqref{twgerbwcon}. In order to define the twisted differential Chern 
character, we proceed very much along the lines of the topological case. For the sake of completeness, we provide a sketch of the proof of 
this construction here. A more detailed account can be found in \cite{GS7}. 

\begin{proposition}[Twisted differential Chern character] \label{prop-twdch}
Let $X$ be a smooth manifold and let $\hat{h}:X\to \BB^2U(1)_{\nabla}$ be a twist for differential K-theory. Let 
$\widehat{\op{K}}_{\hat{h}}\to X$ and $\widehat{H\QQ}[u,u^{-1}]_{\hat{h}}\to X$ be the bundles of sheaves of spectra 
representing $\hat{h}$-twisted differential K-theory and rational differential cohomology (twisted by post-composition
with the canonical map $\BB^2\RR/\ZZ_{\nabla}\to \BB^2\RR/\QQ_{\nabla})$  \cite{GS4}\cite{GS6}. 
There is a morphism of bundles of  sheaves of spectra
$$
\widehat{\ch}_{\hat{h}}:\widehat{\op{K}}_{\hat{h}}\longrightarrow \widehat{H\QQ}[u,u^{-1}]_{\hat{h}}
$$
inducing a twisted differential Chern character map $\widehat{\ch}_{\hat{h}}:\widehat{\op{K}}_{\hat{h}}(X)\to 
\widehat{H}_{\hat{h}}(X;\QQ[u,u^{-1}])$, which locally restricts to the untwisted differential Chern character.
\end{proposition}
\theproof
We again proceed by constructing this map universally. The differential Chern character map 
$\widehat{\ch}:\widehat{\op{K}}\to \widehat{H\QQ}[u,u^{-1}]$
defines a morphism of differential ring spectra (see \cite{BN} or \cite{GS7} for the details of differential ring spectra) 
and hence induces a morphism on the corresponding stack of twists. From the $\infty$-Grothendieck construction, 
this in turn canonically defines a morphism of bundles of sheaves of spectra \cite{GS7}
$$
\xymatrix{
\xi\ar[rr]^-{\widehat{\ch}}\ar[d] &&  \eta\ar[d]
\\
{\rm Tw}(\widehat{\op{K}})\ar[rr]^-{\hat{\ch}} && {\rm Tw}(\widehat{H\QQ}[u,u^{-1}])
}
$$
where the bundles $\xi$ and $\eta$ are the canonical bundles with fiber over an element given by the differential function spectra 
represented by that element. It can be directly verified, using the fact that such a factorization exists in the topological case, 
that precomposition with the map $i:\BB^2U(1)_{\nabla}\to \widehat{\rm Tw}(\widehat{\op{K}})$ factors through a map 
$j:\BB^2\RR/\QQ_{\nabla}\to \widehat{\rm Tw}(\widehat{H\QQ}[u,u^{-1}])$, defined analogously to \eqref{twgerbwcon}. 
This then induces the desired morphism on pullback bundles. That this reduces to the untwisted differential Chern character 
locally follows  by definition and local triviality of the twist.
\endofproof

This completes our discussion of the twisted differential Chern character. The only missing ingredient is a differential refinement of 
the $\hat{A}$-genus, which we now supply.

\begin{remark}[Refinement of the A-genus] {\bf (i)} 
In \cite{GS7}, we defined such a refinement and used it to prove a Riemann-Roch theorem for differential 
$\op{KO}$-theory. Characteristic forms admit unique differential refinements (see \cite{Bun}) and hence there is a natural candidate 
for the $\hat{A}$-genus, given by taking Deligne-Beilinson cup products $\cup_{\rm DB}$ of refined Pontrjagin classes. For a 
Riemannian manifold  $(M,g)$, this leads to a differential cohomology class taking values in $\widehat{H}^*(M;\QQ[u,u^{-1}])$, 
which depends on the metric $g$ and which we denote by $\hat{\mathbb{A}}(M;\nabla_g)$. The first few terms are 
$$
\hat{\mathbb{A}}(M;\nabla_g)=1-\tfrac{1}{24}\hat{p}_1 
+\tfrac{1}{5760}(7\hat{p}_1^2-4\hat{p}_2) +\hdots\;.
$$
Here products such as $\hat{p}_1^2$ mean the Deligne-Beilinson cup product $\hat{p}_1 \cup_{\rm DB} \hat{p}_1$, and so on. 
Taking the formal square root leads to the desired term appearing in \eqref{rrfieltwdf}. 

\item {\bf (ii)} In the formula for the Riemann-Roch theorem in \cite{GS7}, there is a secondary differential form that appears, 
which is related to the $\eta$ form of \cite{BC}. Such forms appear in type IIA string theory in a novel way \cite{MS}
and have been interpretated in terms of the string theory fields \cite{gerbe}. 
In the full differential refinement, it will be important to include this term when taking D-brane charge. 
This will be discussed elsewhere. 
\end{remark}

We now give some indications for why \eqref{rrfieltwdf} is the right definition. There is a pairing in twisted $K$-theory which induces 
Poincar\'e duality. 

\paragraph{\underline{Case 1}: $X$ is a ${\rm Spin}^c$-manifold:} Let $h:X\to K(\ZZ,3)$ be a 
map representing a twist for $K$-theory and let $-h:X\to K(\ZZ,3)$ be the inverse twist. 
We then have a homotopy commutative diagram 
$$
\xymatrix@R=.5em{
 X \; \ar@{^{(}->}[r]^-{\Delta} \ar@/_.7pc/[drr]  &   X\times X\ar[r]^-{(h,-h)} &
 K(\ZZ,3)\times K(\ZZ,3) \ar[r]^-{+} 
 &  K(\ZZ,3)
  \\
&& \ast \ar@/_.5pc/[ur] 
}
$$ 
From the K\"unneth spectral sequence for twisted K-theory \cite{Braun} (see also \cite[Ch. 22]{MSi}),
 we have an induced map
$$
\cup:\op{K}_h(X)\otimes \op{K}_{-h}(X)\longrightarrow \op{K}(X\times X)\xrightarrow{\Delta^*} \op{K}(X)\;.
$$
Postcomposition with the index map $M_{!}:\op{K}(X)\to \op{K}(\ast)\cong \ZZ$, gives a duality pairing. One 
can ask what the cohomological reflection of this map is. For this, we examine the commutativity of the diagram 
$$
\xymatrix{
\op{K}_h(X)\otimes \op{K}_{-h}(X)\ar[r]\ar[d]_-{{\rm ch}_h\otimes {\rm ch}_{-h}} & \op{K}(X)\ar[rr]^-{X_!} 
\ar[d]_-{{\rm ch}} && \op{K}(\ast)\cong \ZZ\ar@{^{(}->}[d]
\\
H_{h}(X;\QQ[u,u^{-1}])\otimes H_{-h}(X;\QQ[u,u^{-1}])\ar[r] & \bigoplus_{i=0}^{\infty}H^{2i}(X,\QQ)
\ar[rr]^-{(X_!)_{d_X}} && H(\ast;\QQ)\cong \QQ
}
$$
where the subscript $d_X$ indicates that we are taking the $d_X={\rm dim}(X)$ component. As is well-known, the right 
square does not commute -- at least, not with the usual Thom isomorphism in cohomology giving rise to the isomorphism 
on the bottom. By the Hirzebruch-Riemann-Roch theorem \cite{AH}, the correction factor to this commutativity is given 
by the $\hat{A}$-genus, twisted by the canonical ${\rm Spin}^c$ line bundle $\mathcal{L}$. Thus, in the special case 
where $X$ has Spin structure, the proposed correction
$
G(x)=\ch_h(x)\cup \sqrt{\hat{A}(X)}
$
indeed makes the diagram commute. In fact, this is essentially the argument originally used in \cite{MM} to deduce the form 
of the anomalous coupling on the worldvolume of $N$ coincident D-branes. Indeed, there the coupling was deduced from 
an anomaly inflow argument (see \cite{CY}). The descent argument used to calculate the anomaly amounts to a simple application of the 
index theorem applied to two transversally intersecting branes with Chan-Paton bundles inherited from the corresponding 
branes. The argument for the square root in the formula is then just that one seeks a pair of forms whose wedge product is 
the calculated anomaly. 

\paragraph{\underline{Case 2}: $X$ does not admit ${\rm Spin}^c$ structure:} In this case one cannot even define the 
pushforward  $X_!$ at the level  of K-theory as this requires  taking the index of the Dirac operator, which is not well-defined. 
This is  precisely the case  that leads to an anomalous action \cite{MW}. However, the \emph{twisted} Thom isomorphism 
allows us to deal even with this case and the above  diagram is modified to 
$$
\xymatrix{
\op{K}_h(X)\otimes \op{K}_{W_3-h}(X)\ar[r]\ar[d]_-{{\rm ch}_h\otimes {\rm ch}_{W_3-h}} & 
\op{K}_{W_3}(X)\ar[rr]^-{X_!} \ar[d]_-{{\rm ch}_{W_3}} && \op{K}(\ast)\cong \ZZ\ar@{^{(}->}[d]
\\
H_{h}(X;\QQ[u,u^{-1}])\otimes H_{-h}(X;\QQ[u,u^{-1}])\ar[r] & 
\bigoplus_{i=0}^{\infty}H^{2i}(X,\QQ)\ar[rr]^-{(X_!)_{d_X}} && H(\ast;\QQ)\cong \QQ
\;,}
$$
where we have used that, rationally, the twist corresponding to $W_3$ vanishes. Note that the 
top composite map can be obtained via the K\"unneth spectral sequence, composed with the 
pushforward for twisted K-theory (see \cite{CW}\cite{CMW}). From the $C^*$-algebra 
point of view this is discussed in \cite{BMRS}.

\medskip
The machinery established in \cite{GS7} and \cite{GS6} allows us to further promote much of this discussion  to the differential case. 
We avoid the case where the twist coincides with a differential refinement of $W_3$, since the twisted Thom isomorphism is no longer 
presented by the tensor product with the virtual spinor bundle, and hence one could argue that the $\hat{A}$ genus should be modified 
from the original formula (we expect that this is not the case though). We will discuss this elsewhere.

\medskip
In \cite{GS7}, we defined the differentially refined $\hat{A}$-genus \footnote{This was done in the context of ${\rm KO}$ theory, but with minor modifications all the arguments hold for $K$-theory -- simply replace ${\rm Spin}$ with ${\rm Spin}^c$ and $\hat{A}$ with $e^{c_1/2}\hat{A}$.} 
and discussed a Riemann-Roch formula. The main theorem asserts that if $f:(X,g)\to (Y,h)$ is a smooth map between manifolds Riemannian 
manifolds, then 
$$
\widehat{\ch}(f_!(E,\nabla))\cup_{\rm DB}\hat{\mathbb{A}}(Y,\nabla_{h})=
\int_{X/Y}\widehat{\ch}(E,\nabla)\cup_{\rm DB} \hat{\mathbb{A}}(X,\nabla_{g+f^*h})+a\big(\ch(\mathcal{F}_{\nabla})\wedge \eta\big),
$$ 
for some odd differential form $\eta$, which is related to the $\eta$-form of \cite{BC}. Given that we have a differential Riemann-Roch 
theorem at our disposal, along with a twisted differential Chern character, the same argument as in the topological case then shows that 
for an RR-field $\hat{x}$ in twisted differential K-theory, its charge $\widehat{G}(\hat{x})$ should be given by expression \eqref{rrfieltwdf}.
This is our proposal for the correct differential cohomological description of the RR fields.

\section{Lifting RR forms} 
\label{Sec-lift} 

\subsection{RR forms arising from differential K-theory}
\label{Sec-liftK}

In what follows, we have in mind ${\rm dim}(X)=10$, although many of the statements hold in greater generality. In 
\cite{MW}, it  was proposed that not just RR charge, but also the RR fields themselves should be regarded as K-theory 
classes. More precisely, in the case where $H_3=0$, we ought to have expression 
\eqref{RR-field ch} for some class $x\in \op{K}(X)$. In \cite{MW}, some simple examples were considered with $\hat{A}(X)=1$ 
which illustrate that, for instance, the fields $G_4$ and $G_6$ are not unrelated, but rather that the class of $G_4$ has an effect 
on the periods of $G_6$ (which are not integral in general). More precisely, if $Sq_2(G_4)=0$ (and $G_2=0$), then $G_6$ 
necessarily has integral periods. If $Sq_2(G_4)\neq 0$, it only has half-integral periods in general. This type of effect is not isolated 
and we will show that our spectral sequence \cite{GS3}\cite{GS5} 
gives a complete list of such conditions which determine when a differential form 
can be lifted to K-theory.
%
%

\medskip
We now turn our attention to the question of which forms $G$ can arise as in \eqref{RR-field ch}. In other words, given a differential 
form $G$, when does it represent an RR-field in twisted differential K-theory. This is equivalent to finding $G^{\prime}$ such that 
$$
G^{\prime}=\frac{G}{\sqrt{\hat{A}(\mathcal{R}_{g})}}={\rm ch}_H(\hat{x})
$$
for some element $\hat{x}\in \widehat{K}_{\hat{h}}(X)$. The formal square root can be calculated as follows. 
For $X$ a 10-dimensional manifold, we have
$$ 
\hat{A}=1-\tfrac{1}{24}p_1 +\tfrac{1}{5760}(7 p_1^2-4p_2) .
$$
Using the formula for the formal square root
$$
\sqrt{\hat{A}}=1+\tfrac{1}{2}\hat{A}_4+\big(\tfrac{1}{2}\hat{A}_8-\tfrac{1}{8}\hat{A}_4^2\big) ,
$$
we can calculate the square root in terms of characteristic forms as
$$
\sqrt{\hat{A}}=1-\tfrac{1}{48}p_1+\big(\tfrac{1}{11520}(7 p_1^2-4p_2)-\tfrac{1}{4608}p_1^2\big) .
$$
Now $\sqrt{\hat{A}}$ is invertible as a differential form (since it is of the form $1+x$ with $x$ nillpotent). Therefore, a differential form $G=G_0+G_2+\hdots $ is always in the image of the map $\sqrt{\hat{A}}\wedge$, with $G^{\prime}=\frac{G}{\sqrt{\hat{A}}}$ mapping to $G$. Using the formula
$(1+x)^{-1}=1-x+x^2-x^3+\hdots$,  we immediately calculate 
$$
\big(\sqrt{\hat{A}}\big)^{-1}=1+\tfrac{1}{48}p_1-\big(\tfrac{1}{11520}(7 p_1^2-4p_2)
-\tfrac{1}{4608}p_1^2\big)+\tfrac{1}{2304}p_1^2
=1+\tfrac{1}{48}p_1-
\big(\tfrac{1}{11520}(7 p_1^2-4p_2)-\tfrac{3}{4608}p_1^2\big).
$$
The condition then becomes that the formal power series
\bea
G^{\prime}=\frac{G}{\sqrt{\hat{A}}} &=& G_0
\\
&+& G_2
\\
&+& \big(G_4+\tfrac{1}{48}p_1 G_0\big)
\\
&+& \big(G_6+\tfrac{1}{48}p_1 \wedge G_2\big)
\\
&+& \big(G_8+\tfrac{1}{48}p_1 \wedge G_4- \tfrac{1}{11520}(7 p_1^2-4p_2)-
\tfrac{3}{4608}p_1^2\big)\wedge G_0\big)
\\
&+&  \Big(G_{10}+\tfrac{1}{48}p_1\wedge G_6-
\big(\tfrac{1}{11520}(7 p_1^2-4p_2)
-\tfrac{3}{4608}p_1^2\big)\wedge G_2\Big)
\eea
is in the image of the Chern character map. The goal for the remainder of this section will be to calculate the image of the Chern 
character purely in terms of conditions in cohomology. In other, words, we seek necessary and sufficient conditions on $G^{\prime}$ 
so that its components lift through the Chern character to differential K-theory. To do this, we will utilize the AHSS developed 
in \cite{GS3}. We first review the necessary material.

\medskip
The AHSS for topological K-theory has $E_2$-page whose groups are given by cohomology with integral coefficients appearing 
periodically in the the degree indexing $\pi_q(\op{K})$. More precisely, we have $E_2^{p,q}=0$ is $q$ is odd and 
$E_2^{p,q}=H^p(X;\ZZ)$ if $q$ is even. By degree considerations $d_2=0$ and the the first nonvanishing 
differential $d_3$ is given by the  formula \cite{Ro}\cite{AS2} 
$$
\xymatrix{
d_3=Sq^3_\Z : H^p(X; \Z)
\; \ar[r] &
H^{p+3}(X; \Z)\;,
 }
$$
where $Sq^3_{\ZZ}$ is the composite operation $\beta Sq^2\rho_2$, and $\beta:H^p(X;\ZZ/2)\to H^{p+1}(X;\ZZ)$ is the 
Bockstein homomorphism associated to the mod 2 reduction sequence. 

\medskip
The AHSS for differential K-theory \cite{GS3}\cite{GS5} shares some similarities with its topological counterpart with two crucial distinctions. 
The $E^{0,0}_2$-entry of the spectral sequence is singled out as containing the geometric information given by differential forms. 
In fact, $E^{0,0}_2=\Omega^{\rm even}_{\ZZ,{\rm cl}}(X)$ whose elements are formal combinations of closed differential forms 
$\omega=\omega_0+\omega_2+\hdots$ with $\omega_0\in \ZZ$ and $ \omega_p $ has degree $p$. There is also a difference in 
coefficients for the other terms on the $E_2$-page (from $\ZZ$ to $U(1)$) and a shift in the degree indexed by $q$. This shift is 
essentially due to the shift from the Bockstein homomorphism $H^p(X;U(1))\to H^{p+1}(X;\ZZ)$ associated to the exponential 
sequence and the permanent cycles at these stages correspond to the torsion information in differential K-theory. Summarizing, 
the $E_2$-page looks as follows:

{\footnotesize
\(
\label{E2-Kth}
\begin{tikzpicture}
  \matrix (m) [matrix of math nodes,
    nodes in empty cells,nodes={minimum width=5ex,
    minimum height=4ex,outer sep=-5pt}, 
    column sep=1ex,row sep=1ex]{
                &      &     &     & 
                \\         
            1 &     &   &   &
            \\
             0     &   \Omega^{\rm even}_{\ZZ,{\rm cl}}(X)&  &   &
             \\           
               -1\  \   &  & H^1(X;U(1)) &  H^2(X;U(1))     & 
                \\
              -2\ \  &   &    &  0 & 0 & 0
               \\
               -3\ \ &  & & &  & H^4(X;U(1))
               \\
               -4\ \ &   &    &   &  &  & 
               \\
                &      &     &     & \\};
                
  \draw[-stealth] (m-3-2.east) -- node[above]{\small $d_2$} (m-4-4.north west);
   \draw[-stealth] (m-4-3.south east) -- (m-5-5.west);
      \draw[-stealth] (m-5-4.south east) -- (m-6-6.north west);
        \draw[-stealth] (m-4-4.south east) -- (m-5-6.north west);
\draw[thick] (m-1-1.east) -- (m-7-1.east) ;
\end{tikzpicture}
\)
}
while the $E_3$-page looks as
{\footnotesize
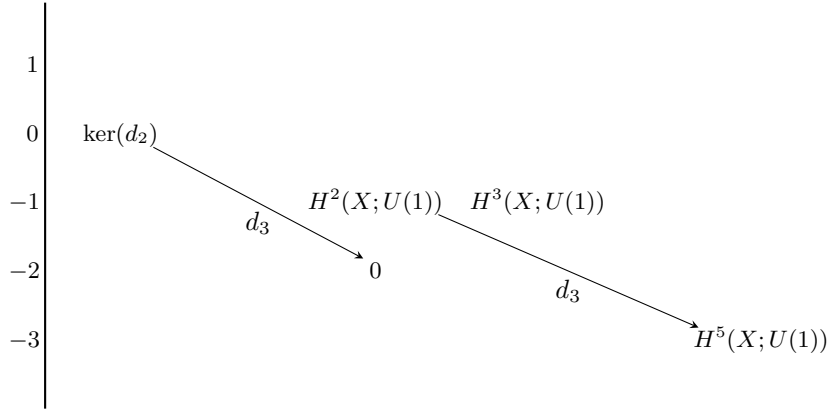
$$
\begin{tikzpicture}
  \matrix (m) [matrix of math nodes,
    nodes in empty cells,nodes={minimum width=5ex,
    minimum height=5ex,outer sep=-5pt}, 
    column sep=1ex,row sep=1ex]{
                &      &     &     & 
                \\         
            1 &     &   &   &
            \\
             0     &   \ker(d_2)  &  &   &
                \\
              -1\ \  &   &    &  & H^2(X;U(1))  & H^3(X;U(1)) 
               \\
               -2\ \ &  &  &  & 0 
               \\
               -3\ \ &   &    &   &  &  & & H^5(X;U(1))
               \\
                 &      &     &     & \\};
                 
             \draw[-stealth] (m-3-2.south east) --node[pos=.5,below]{\small $d_3$} (m-5-5.north west);      
   \draw[-stealth] (m-4-5.south east) --node[pos=.5,below]{\small $d_3$} (m-6-8.north west);
\draw[thick] (m-1-1.east) -- (m-7-1.east) ;
\end{tikzpicture}
$$
}
This pattern continues, taking kernel mod image at each stage. From \cite[Proposition 26]{GS3}, we have the identification of the differentials 
in the lower quadrant $q<0$ as 
$$
\widehat{Sq}^3=jSq^2\rho_2\beta_{U(1)}:H^p(X;U(1))\longrightarrow H^{p+3}(X;U(1)),
$$
where $j:H^p(X;\ZZ/2)\to H^p(X;U(1))$ is the inclusion as the 2-roots of unity and $\beta_{U(1)}$ is the Bockstein associated to the 
exponential sequence. The hat notation for the above operation is justified by the fact that this operation is the restriction of the only 
natural operation in differential cohomology which refines the second Steenrod square \cite{GS2}. In general, we have the following 
two types of differentials in the spectral sequence \cite{GS3}
\begin{enumerate}[label=(\roman*)]

\vspace{-2mm}
\item \emph{(Obstructions associated to curvature forms):} Those give
 obstructions for the curvature forms to lift to differential $K$-theory and are of the form
\(\label{type 1 diff}
d^{0,0}_{2k}:E^{0,0}_{2k-1}\subset 
\Omega^{\rm even}_{\ZZ,{\rm cl}}(X) \longrightarrow
 E^{2k,2k-1}_{2k-1}\subset H^{2k}(X;U(1))/{\rm im}(d_{p-1}),
\)
where $E^{0,0}_{2k-1}=\bigcap {\rm ker}(d_{2j})$ for $k<j$.

\vspace{-2mm}
\item \emph{(Obstructions associated to flat classes):} These 
give rise to obstructions to lifting to flat classes (see e.g. \cite{Lo} \cite{Ho2} for 
a description of such classes):
\(\label{type 2 diff}
d^{p,-2q+1}_{2k+1}:H^{p}(X;U(1)) \longrightarrow H^{p+2k+1}(X;U(1)),
\)
emanating from the entries $E^{p,-2q+1}_{2k+1}$, with $q\geq 0$. 
\end{enumerate} 

We will need identifications of both types of differentials to get a good understanding of the spectral sequence. We now proceed 
with this identification. In the topological case, it was shown in \cite{Buc1}\cite{Buc2} that we have the following differentials 
in the AHSS for  topological K-theory restricted to the $p$-primary part of $H^*(X;\ZZ)$. The notation $d^p_n$ denotes 
the $p$-primary part  of the differential $d_n$:
\begin{align}\label{buchdf}
d_3(x) &=\beta_2 Sq^2(x) 
\\
d_5^2(2x) &=\beta_2 Sq^4(x), \qquad d_5^3(x) = \beta_3 \mathscr{P}^1_3(x)\;, 
\nonumber
\\
d_7^2(4x) &= \beta_2 Sq^6(x) \nonumber
\\
d_9^2(8x) &=  \beta_2 Sq^8(x), \qquad   d_9^3(3x) =  \eta_3\beta_3 \mathscr{P}_3^2(x)\;, \qquad  d_9^5(x) = 
 \eta_5\beta_3 \mathscr{P}_5^1(x)\;, \nonumber
\end{align}
with $\eta_p\neq 0 \mod p$. Some of these differentials have appeared in studying anomalies in M-theory in \cite{KSpin}. 
Now we have the following in the differential refinement, which identifies a portion of the 
differentials of type \eqref{type 2 diff}.
\begin{proposition}[Torsion differentials in the $\widehat{\rm AHSS}$]
\label{dfrefbu}
We have the following $p$-primary parts of the differentials in the AHSS for $\widehat{K}$, occurring in the lower quadrant of the 
half-plane spectral sequence for $\widehat{K}$. 
\begin{align}
d_3(x) &=j_2 Sq^2\rho_2\beta_{U(1)}(x)  \nonumber
\\
d_5^2(2x) &=j_2Sq^4\rho_2\beta_{U(1)}(x), \qquad  d_5^3(x) = j_3 \mathscr{P}^1_3\rho_3\beta _{U(1)}(x), \nonumber
\\
d_7^2(4x) &= j_2 Sq^6\rho_2\beta_{U(1)}(x) \nonumber
\\
d_9^2(8x) &=  j_2Sq^8\rho_2\beta_{U(1)}(x), \qquad  d_9^3(3x) =  \eta_3j_3\rho_3 \mathscr{P}_3^2\beta_{U(1)}(x), \qquad 
 d_9^5(x) =  \eta_5\rho_5j_5\mathscr{P}_5^1\beta_{U(1)}(x), \nonumber
\end{align}
where $\eta_p\neq 0 \mod p$ and $j_p:H^k(X;\ZZ/p)\into H^k(X;U(1))$ is the induced by the inclusion as the primitive 
$p$-roots of unity.
\end{proposition}
\theproof
The connecting map $K^{*-1}_{U(1)}(X)\to K^*(X)$ in the exponential exact sequence for $K$-theory with coefficients induces 
a morphism of spectral sequences. This morphism vanishes by degree considerations. In such a situation, there is a well-defined 
boundary morphism of spectral sequences and the argument in \cite{GS3} shows that the Bockstein homomorphism in cohomology 
$\beta_{U(1)}:H^*(X;U(1))\to H^{*+1}(X;\ZZ)$ commutes with the differentials in the AHSS. From the identification
 \eqref{buchdf} and the relation $\beta_{U(1)}j_p=\beta_p$, the claim follows.
\endofproof

Since $\mathscr{P}^n(x)=0$ for $2n>{\rm deg}(x)$, $Sq^n(x)=0$ for $n>{\rm deg}(x)$ and $H^k(X;U(1))=0$ for $k>10$, 
we see that there is only a total of only three differentials in Proposition \ref{dfrefbu} which need not vanish. These are 
\begin{align}
d_3(x) &=j_2 Sq^2\rho_2\beta_{U(1)}(x)  \nonumber
\\
d_5^2(2x) &=j_2Sq^4\rho_2\beta_{U(1)}(x) , \qquad  d_5^3(x) = j_3 \mathscr{P}^1_3\rho_3\beta _{U(1)}(x) . \nonumber
\end{align}
At the level of cohomology, these act by 
$$
\xymatrix@C=.2cm{
H^0 & H^1\ar@/^2pc/[rrr]^-{j_2 Sq^2\rho_2\beta_{U(1)}}\ar@/_2pc/[rrrrr]_<<<<<<<<<{j_3 \mathscr{P}^1_3\rho_3\beta_{U(1)}}  
& H^2& H^3\ar@/_2pc/[rrrrr]_>>>>>>>>{j_2Sq^4\rho_2\beta_{U(1)}} & H^4 & H^5 & H^6 & H^7 & H^8 & H^9 & H^{10}
}
$$
where the arrows indicate the first nonvanishing occurrence of the operations, and other operations are right-translates. 
The differentials identified above provide information on the torsion in $\widehat{K}$ and also constrain the permanent 
cycles in $E^{0,0}_{\infty}$. However, in order to have a truly satisfactory condition for differential forms to lift to 
$\widehat{K}$, we really need to identify the differentials of type \eqref{type 1 diff}. Since these differentials are completely 
responsible for calculating the image of the Chern character, it is not surprising that they are very rich and combine the geometry 
and topology of the manifold in a non-trivial way. As a consequence, the formulas for the differentials can be difficult to parse, 
and some explanation leading up to the identification is in order. 

\medskip
For a prime $p$, we can speak of the $p$-primary part of the differential $(d^{0,0}_n)^p$ as the part of the differential which 
factors through the inclusion $H^n(X;\ZZ/p)\into H^n(X;U(1))$ as the primitive $p$-roots of unity. By convention, we take 
$p=0$ to be the part of the differential which factors through the exponential ${\rm exp}:H^n(X;\RR)\to H^n(X;U(1))$. 
The primes $p$ will be responsible for shifted quantization  condition of the form
$$
\int_{C}G_{2k}=\frac{1}{p^r}\int_C\lambda+{\rm integer}\;,
$$
with $\lambda$ some differential form, $r>1$ an integer and $C$ a cycle in spacetime. As we show below, the differentials take the form 
$$
d=\text{Cohomology operation} + \text{Exponential of differential form} .
$$
The cohomology operations in the formula are only defined on the kernel of the previous differentials and are a combination of Steenrod 
squares and powers. These terms contribute to the $p$-primary part of $d$, with $p\neq \infty$ and thus are responsible for shifting the 
usual quantization. We are now ready to identify the differentials. In what follows, we only consider the differentials \eqref{type 1 diff} 
and we drop the superscript indicating the bidegree.

\begin{proposition} [Degree two]\label{iddifd2}
The differential $d_2$ is given by 
$$
d_2(\omega)=[\omega_2]\!\!\!\mod \ZZ .
$$
\end{proposition}
\theproof
First observe that every exact form lifts to $\widehat{K}$. Indeed, as part of the data of the differential refinement, there is a 
canonical map $a:\Omega^{\rm odd}(X)/{\rm im}(d)\to \widehat{K}$ which makes the diagram
$$
\xymatrix@R=1.5em{
\Omega^{\rm odd}(X)/{\rm im}(d)\ar[rr]^-{d}\ar[dr]_-{a} && \Omega^{\rm even}_{\rm cl}(X)
\\
 &\widehat{K}(X)\ar[ru]_-{\mathcal{R}} &
}
$$
commute. Thus, for any exact even form $\omega$ with global potential $d\eta=\omega$, the class $a(\eta)$ defines a refinement of 
$\omega$. Thus, the differential $d_2$ factors through $H^{\rm even}(X;\RR)$ and thus defined a natural transformation of functors 
$d_2:H^{\rm even}(-;\RR)\to H^2(-;U(1))$. 

Now $H^{\rm even}(-;\RR)$ is representable by the product of Eilenberg-MacLane spaces $\prod_{n\in \NN}K(\RR;2n)$ and 
standard arguments in homotopy theory show that the $d_2$ must be the projection onto $H^2$ followed by a map 
$\lambda: H^2(X;\RR)\to H^2(X;U(1))$, where $\lambda:x\to \lambda x\mod \ZZ$, with $\lambda\in \RR$. From the 
identification $\ch_1=c_1$, an integral class, one immediately sees that $\lambda=1$ and $d_2$ is as claimed.
\endofproof

\begin{proposition}[Degree four] \label{d4iddf}
The differential $d_4$ is given by 
$$
d_4(\omega)=([\omega_4]\!\!\!\mod \ZZ)+j_2Sq^2\rho_2(x_2)\;,
$$
where $x_2$ is an integral class representing $\omega_2\in \ker(d_2)=H^2(X;\ZZ)\cap H^2(X;\RR)$.

\end{proposition}
\theproof
First observe that the differential $d_4$ is not only natural with respect to smooth maps $f:X\to Y$ between manifolds, but in fact 
$d_4(f^*\omega)$ only depends on the underlying homotopy class of $f$. Indeed, since exact forms are necessarily killed by the 
differential, it must factors through the cohomology group $H^{\rm even\geq 2}(X;\RR)$. A straightforward argument similar to 
that in Proposition \ref{iddifd2} shows that the restriction of $d_4$ to $H^{\rm even\geq 4}(X;\RR)$ is sends 
$\omega\mapsto [\omega_4]\mod \ZZ$. We need only identify the restriction to $\ker(d_2)\cap H^2(X;\RR)$. 

The proof proceeds by considering the following universal example. Consider the fiber sequence
$$
\xymatrix{
K(U(1),1)\ar[r]^-{\beta_{U(1)}} & K(\ZZ,2)\ar[d] &
\\
& K(\RR,2)\ar[r]^-{\rm exp} & K(U(1),2)
}
$$
and take as a model $\CC P^{\infty}\simeq K(\ZZ,2)$. Now if $M$ is any compact smooth manifold, then a map 
$M\to \CC P^{\infty}$  factors (up to homotopy) through some $\CC P^N\into \CC P^{\infty}$, for $N$ large. 
By the universal property of the homotopy fiber  and the identification of the differential in Proposition \ref{iddifd2}, 
it suffices to prove the claim for each $\CC P^N$. Since  $H^*(\CC P^N;\ZZ)$ is torsion free, it follows that $d_3$ 
vanishes and $\ker(d_2)\cap H^2(\CC P^N;\RR)=H^2(\CC P^N;\ZZ)\cong \ZZ$,  generated by $c_1$. Now $d_4$ 
cannot vanish on $H^2(\CC P^N;\ZZ)$ since this would imply, by the identification of the restriction of  $d_4$ to $H^4$, 
that every degree $4$ component of a Chern character on $\CC P^N$ has integral periods, which is not true 
(i.e., ${\rm ch}_2(\mathcal{L})=\tfrac{1}{2}c_1^2(\mathcal{L})$ with $\mathcal{L}\to \CC P^N$ the canonical line bundle). 
The only other possibility is  homotopy class of the map 
$$
\xymatrix{
\CC P^N \; \ar@{^{(}->}[r]^-{c_1} & 
\CC P^{\infty}\simeq K(\ZZ,2)\ar[r]^-{Sq^2\rho_2} &
 K(\ZZ/2,4)\ar[r]^-{j_2} & K(U(1),4)
}
$$
and, therefore, $d_4$ is as claimed. 
\endofproof

\begin{proposition}[Degree six]
\label{prop-deg6}
The differential $d_6$ is given by 
$$
d_6(\omega)=([\omega_6]\!\!\!\mod \ZZ) +j_3\mathscr{P}_3^1(x_2)+j_2(\overline{Sq}^2)(\omega_4) ,
$$
where $j_2(\overline{Sq}^2)$ is a natural operation, well-defined modulo the image of $d_3$, which restricts on the classes 
$\omega_4\in H^4(X;\ZZ)\cap H^4(X;\RR)$ to $j_2Sq^2\rho_2(x_4)$ with $x_4$ an integral lift of $\omega_4$.
\end{proposition}
\theproof
As before, it is straightforward to show that the restriction of $d_6$ to $H^{\rm even\leq 6}(X;\RR)$ sends 
$\omega\mapsto [\omega_6]\mod \ZZ$. We focus our attention on the restriction to $\ker(d_4)\cap H^2(X;\RR)\oplus H^4(X;\RR)$. 
Again, we proceed by universal example. Consider the 2-stage Postnikov tower
$$
\xymatrix{
K(U(1),3)\ar[r] & X_2\ar[d] & 
\\
 & K(\ZZ,2)\times K(\RR,4)\ar[r]^-{d_4} & K(U(1),4)
 }
$$
with $d_4$ identified as in Proposition \ref{d4iddf}. The $E_2$-page of the Serre spectral sequence for the above 
fibration (with $U(1)$-coefficients) can be identified in the relevant part as follows
$$
\xymatrix@C=.1cm@R=.1cm{
&&&
\\
& j_2Sq^2\rho_2\beta_{U(1)} & 0 &  j_2Sq^2\rho_2\beta_{U(1)} \cdot u& & & & 
\\
& 0 & 0 & 0 &  &  & & 
\\
& 0 & 0 & 0 & 0 &  & & 
\\
& 0 & 0 & 0 & 0 & 0& & 
\\
& 0 & 0 & 0 & 0 & 0 & 0&
\\
& 1 & 0 & u & 0 & (v , j_2Sq^2\rho_2) & j_3\mathscr{P}^1\rho_3 & 0
\\
\ar[uuuuuuu] \ar[rrrrrrr] &&& &&&&
}
$$
with $u$ generating $H^2(K(\ZZ,2);\RR)$ as a vector space, hence $H^2(K(\ZZ,2);U(1))$ modulo $\ZZ$, and $v$ 
generating $H^4(K(\RR,4);U(1))$. The terms on the diagonal with 
bidegrees $(p,6-p)$ converge to $H^6(X_2;U(1))$ and clearly both $ j_2Sq^2\rho_2\beta_{U(1)} $ and 
$j_2Sq^2\rho_2\beta_{U(1)} $ survive to the $E_{\infty}$-page. We conclude that 
$H^6(X_2;U(1))$ is generated by $p^*j_3\mathscr{P}^1\rho_3$ and $j_2\overline{Sq}^2$ with 
$i^*j_2\overline{Sq}^2=j_2Sq^2\rho_2\beta_{U(1)}$.
Through a sequence of surgeries, we can approximate $X_2$ be a sufficiently connected map $f:M\to X_2$, with $M$ a finite-dimensional 
smooth manifold. Furthermore, since $X_2$ represents the universal space for which $d_2$ and $d_4$ vanish, it follows that 
\(\label{d6dfwld}
d_6(\omega)=([\omega_6]\!\!\!\mod \ZZ)+\lambda j_3\mathscr{P}^1(x_2)+\delta j_2(\overline{Sq}^2)(\omega_4)
\)
with $\lambda=0,1,2$ and $\delta=0,1$.

It remains only to show that the restriction to $H^4(X;\ZZ)\cap H^4(X;\RR)$ agrees with $j_2Sq^2\rho_2$ and that $\lambda=\delta=1$. 
For the former, observe that the component $d_4(\omega)=[\omega_4]\mod \ZZ$ vanishes on this restriction.  Hence, for such classes 
we can restrict to the fiber $X^{\prime}_2$ of $j_2Sq^2\rho_2:K(\ZZ,2)\to K(U(1),4)$. Again computing via the Serre spectral sequence 
one easily sees that $k^*(j_2\overline{Sq}^2)=j_2Sq^2\rho_2$, where $k:X^{\prime}_2\to X_2$ is the canonical map. To show that 
$\lambda=\delta=1$, it suffices to consider the example $\CC P^N$. The canonical line bundle $\mathcal{L}\to \CC P^N$ has 
$\ch_3(\mathcal{L})=\tfrac{1}{3!}c^3_1(\mathcal{L})$ and since $c_1^3$ generates $H^6$, the vanishing condition 
$d_6\ch_3(\mathcal{L})=0$ and equation \eqref{d6dfwld} forces us to have 
$$
2\lambda+3\delta\equiv -1 \!\!\!\mod 6.
$$
Hence, $\delta\equiv 1\mod 2$ and $\lambda\equiv 1\mod 3$.
\endofproof

The following condition gives the shifted quantization law for $G_6$ in the general case.

\begin{corollary}[Shifted quantization for $G_6$] 
On an arbitrary manifold $X$, an RR-field $G_6$ necessarily has periods in $\frac{1}{6}\ZZ$. 
\item {\bf (i)} If $\overline{Sq}^2$ 
vanishes on $X$, then $G_6$ has periods in $\frac{1}{3}\ZZ$.

\item {\bf{(ii)}}  If $\mathscr{P}_3^1=0$, then $G_6$ has half integral periods. 

\item {\bf{(iii)}}  If both $\overline{Sq}^2$ 
and $\mathscr{P}_3^1$ vanish, then $G_6$ has integral periods. 
\end{corollary}

We provide illustrations of the above with the following examples. 

\begin{example}[Even spheres]\label{Ex-S2n}
Let $X=S^{2n}$. Then the only relevant nontrivial differential on forms in the AHSS is 
$$
d_{2n}:\Omega^{\rm even}_{\ZZ,{\rm cl}}(S^{2n})\longrightarrow H^{2n}(S^{2n};U(1))\cong U(1) .
$$
Hence $\ch_{n}$ is the component of a Chern-character if and only if $\ch_n$ has integral periods. 
Hence, in particular, for any line  bundle $\tfrac{1}{n!}c_1^n$ is integral. 
\end{example}

\begin{example}[Complex projective spaces]\label{Ex-cpn}
Let $X=\CC P^n$. The first Chern class of the canonical line bundle $\mathcal{L}\to \CC P^n$ 
generates $H^2(\CC P^n;\ZZ)$  and, moreover, $c_1^n$ generates $H^{2n}(\CC P^n;\ZZ)$. 
Thus, the degree $2k$ component of the Chern character
$$
\ch(\mathcal{L})=1+c_1+\tfrac{1}{2}c_1^2 +\hdots + \tfrac{1}{n!}c_1^n\;,
$$
does not represent an integral class and must have periods in $\tfrac{1}{k!}$. From the general 
formula of the differential in Proposition \ref{d4iddf} and the basic properties of the Steenrod algebra, 
we have that the condition for vanishing of the differential $d_4$ is 
$$
j_2(\overline{c}_1^2) \equiv [\ch_2]\!\!\mod \ZZ .
$$
Hence, in particular $\ch_2$ has only half integral periods. This is consistent with the fact that for the canonical 
line bundle $\mathcal{L}$, the class $\tfrac{1}{2}c_1^2(\mathcal{L})$ has only 
half integral periods. The general vanishing condition on $d_6$ is
$$
-j_3(\overline{c}_1^3)+j_2\big(\overline{Sq}^2\big(\tfrac{1}{2}c_1^2\big)\big)\equiv \ch_3\!\!\mod \ZZ .
$$
and so the periods of $\ch_3$ must lie in $\frac{1}{6}\ZZ$ in general. In the particular case of ${\rm ch}(\mathcal{L})$, 
we have \footnote{Note that we are denoting the abelian group operation on $U(1)$ by $+$ on the left. When writing 
these classes in terms of exponentials, we will denote the group operation by juxtaposition, or $\cdot$, identifying it 
with multiplication of complex numbers.}
\(\label{chlbun}
-j_3(\overline{c}_1^3)+j_2\big(\overline{Sq}^2\big(\tfrac{1}{2}c_1^2\big)\big)={\rm exp}(\tfrac{1}{6}c_1^3)\;.
\)
From the right commutative diagram 
\(\label{comdg23}
\xymatrix{
H^*(X;\ZZ)\ar[r]^{\times \frac{1}{2}}\ar[d]_-{\rho_2} &H^*(X;\RR)\ar[d]^-{\rm exp}
\\
H^*(X;\ZZ/2)\ar[r]^{j_2} & H^*(X;U(1))
} 
\qquad \qquad  \qquad 
\xymatrix{
H^*(X;\ZZ)\ar[r]^{\times \frac{1}{3}}\ar[d]_-{\rho_3} &H^*(X;\RR)\ar[d]^-{\rm exp}
\\
H^*(X;\ZZ/3)\ar[r]^{j_3} & H^*(X;U(1))
} 
\)
it follows that $j_3(\overline{c}_1^3)={\rm exp}(\frac{1}{3}c_1^3)$. Plugging this into equation \eqref{chlbun} 
and using the left commutative diagram in \eqref{comdg23}, we have
$$
j_2\big(\overline{Sq}^2\big(\tfrac{1}{2}c_1^2\big)\big)=j_2(\overline{c}_1^3)\;.
$$
It is interesting to compare this with the formula for the Steenrod squares on powers of $c_1(\mathcal{L})$
$$
Sq^{2r}(\overline{c}_1^n)={n\choose r}\overline{c}_1^{n+r} \quad \Rightarrow  \quad 
Sq^2(\overline{c}_1^2)=2\overline{c}_1^3=0.
$$

\end{example}

\begin{example}[${\rm Spin}^c$ 4-manifolds]
Let $M$ be a 4-dimensional manifold with ${\rm Spin}^c$ structure and let $\lambda=c_1(\mathcal{L})$, with $\mathcal{L}\to M$ 
the canonical line bundle associated to the structure. Let $E$ be any complex vector bundle on $M$. Then 
$$
\ch(E\otimes\mathcal{L})=\ch(E)\wedge \ch(\mathcal{L})=\ch(E)+(r+c_1(E))\lambda
$$
with $r\in \ZZ$ the rank of $E\to M$. Hence
$$
\ch(E\otimes\mathcal{L}-E)= (r+c_1(E))\lambda\;.
$$
Now $\ch(E\otimes\mathcal{L}-E)_4$ must be killed by $d_4$, since it is the Chern character of a virtual bundle. 
Hence, we must have 
$$
c_1(E)\lambda \equiv j_2Sq^2\rho_2 c_1(E) \mod \ZZ ,
$$
which recovers the well-known relation $\nu_2=w_2=\lambda \mod 2$. Clearly, $\lambda$ depends on the choice of 
${\rm Spin}^c$ structure. It defines a characteristic element of the bilinear pairing on $H^2(M; \Z)$, 
defined by the cup product, i.e.,
$$
\int_M c_1(E)^2\equiv \int_Mc_1(E)\lambda \mod 2 .
$$
The associated quadratic refinement of the intersection pairing has been studied in many places,  for instance Atiyah \cite{At} 
in his work  on Riemann surfaces. In \cite{HS}, a higher dimensional analogue of this pairing was used in the construction of the 
fivebrane partition function. 
\end{example}

\begin{remark}[Novel quantization conditions] 
Interestingly, the machinery of our AHSS provides a way to determine the quantization condition on the fields $G_{2k}$ purely 
in cohomology. This addresses a key point made in \cite{MW}, where it is assumed that such conditions would be nearly impossible 
to determine purely in cohomology. While this certainly seems to be the case without any reference to K-theory, the spectral sequence 
uses the differential K-theoretic interpretation as a starting point and interprets the quantization conditions in cohomology as an 
obstruction to lifting a form to differential K-theory. The differentials in the AHSS precisely measure the obstruction to lifting.
\end{remark}


The following is then a direct consequence of the identification of the differentials in the AHSS.

\medskip
\begin{proposition}[Algorithm for detecting RR-fields] Let 
$$
G=G_0+G_2+G_4+G_6+G_8+G_{10}
$$
be a formal combination of forms on spacetime $X$. Then the following provide necessary and sufficient conditions 
on the components $G_{2k}$, with $k\leq 3$ so that $G_{2k}$ lifts to differential K-theory.
\begin{enumerate}
\vspace{-1mm} 
\item For $G^{\prime}_0=G_0$, we have $G_0\in \ZZ$.
\vspace{-1mm} 
\item For $G^{\prime}_2=G_2$, we have the condition $[G_2]\equiv 0 \mod \ZZ$ so that $G_2$ has integral periods. 
\vspace{-1mm} 
\item 
For $G^{\prime}_4=G_4+\tfrac{1}{48}p_1G_0$, we must have 
$$
([G^{\prime}_4] \!\!\!\mod \ZZ) = j_2Sq^2\rho_2(x_2) 
$$
for some class $x_2\in H^2(X;\ZZ)$ which defines an integral lift of $[G_2]$.
\vspace{-1mm} 
\item For $G^{\prime}_6=G_6+\tfrac{1}{48}p_1\wedge G_2$, we must have
$$
([G^{\prime}_6] \!\!\!\mod \ZZ) = j_2\overline{Sq}^2(x_4) - j_3 P^1_3\rho_3(x_2) 
$$
for some $x_4\in \ker(d_4)\oplus {\rm Tor}(H^4(X; \Z))$ and $x_2\in H^2(X;\ZZ)$, where the 
$x_4$ and $x_2$ rationalize to $[G_4]$ and $[G_2]$, respectively. 
 In particular, it is sufficient that 
$$
([G^{\prime}_6] \!\!\!\mod \ZZ) =j_2Sq^2(x_4) - j_3 P^1_3\rho_3(x_2) 
$$
with $x_4\in H^4(X; \Z)$ and $x_2\in H^2(X; \Z)$. 
\end{enumerate}
\end{proposition}

The algorithm can in principle be extended to $G_8$ and $G_{10}$, but the expressions 
would become very complicated.

\subsection{RR forms arising from twisted differential K-theory}
\label{Sec-liftKt}

We now consider the twisted case. 
The first differential $d_3$ in the twisted AHSS for twisted K-theory 
is given by the  formula \cite{Ro}\cite{AS2}
$$
\xymatrix{
d_3=Sq^3_\Z + (-) \cup  \lambda h: H^p(X; \Z)
\; \ar[r] &
H^{p+3}(X; \Z)\;,
 }
$$
where $\lambda$ is an integer which a priori needed to be determined. To compute this integer, it is sufficient to consider 
the spectral sequence on the sphere $S^3$, where one computes $\lambda=-1$ (see \cite{AS2}). To our knowledge, this is 
the only differential which is identified explicitly in the twisted case. However, Atiyah and Segal \cite{AS2} also showed that
 the higher differentials $d_5, d_7, \cdots$ in the AHSS for twisted K-theory are nontrivial even rationally, and are given by Massey products. 
In order to work with smooth manifolds, it is easier to take real coefficients, i.e., work 
over $\RR$, in which case differential forms can be used as chains. 

\medskip
Working with twisted K-theory over $\RR$, i.e. essentially 
periodic twisted cohomology, the iterated
Massey products with the twist $H_3$ gives (up to sign) all the 
higher differentials in the tAHSS for twisted cohomology \cite{AS2}
$$
d_{2i+1}(x)=-\langle \; \underbrace{[H_3], \cdots, [H_3]}_{i~ {\rm times}}, x \rangle\;. 
$$
\begin{enumerate}
\item The class in the $E_4$-page is given by 
the triple Massey product $\langle H_3, H_3, x_n \rangle$, where $H_3$ is the twisting 
cohomology class and $x_n$ is the dimension $n$ class under consideration. 
Since $H_3\wedge H_3=0$, then
$$
\langle H_3, H_3, x_n \rangle= y_{n+2}\wedge H_3\;,
$$
where 
$
H_3 \wedge x_n= dy_{n+2}
$.
This operation corresponds to the differential $d_5: E_4^p \to E_4^{p+5}$.

\vspace{-2mm}
\item Next, when $\langle H_3, H_3, x_n\rangle =0$, i.e.,
$\langle H_3, H_3, x_n\rangle = dz_{n+4}$ modulo multiples of $H_3$, then 
 the next step gives the quadruple Massey product
 $$
 \langle H_3, H_3, H_3, x_n \rangle = H_3 \wedge z_{n+4}\;,
 $$
which corresponds to the differential $d_7: E_6^p \to E_6^{p+7}$ on the 
$E_6$-page. 
\end{enumerate}

\begin{example}[Dynamics of twisted RR fields via Massey products] 
We consider the Ramond-Ramond (RR) fields $F_i$,
twisted by the NS field $H_3$. We start with a class corresponding to a specific 
degree, so that $x_n$ is identified with the class of $F_n$, and we will use the 
latter as notation. Then we have 
$$
\langle H_3, H_3, F_n \rangle= F_{n+2}\wedge H_3 
$$
where 
\(
\label{Htwcl}
H_3 \wedge F_n= dF_{n+2}\;,
\)
which is the correct equation of motion/Bianchi identity for the fields.
This is the differential $d_5$ in the twisted AHSS. 
Note that because $H_3$ is closed odd form, and due to equation 
\eqref{Htwcl}, we have the 
closedness of the Massey triple product, i.e., $d \langle H_3, H_3, F_n \rangle=0$.
Next, if we trivialize the triple Massey product, i.e., take 
$F_{n+2}\wedge H_3=dF_{n+4}$, which is the correct dynamics in the 
next level up in RR degrees, then we can form the quadruple Massey
product 
\(
\label{nextH}
\langle H_3, H_3, H_3, F_n \rangle= F_{n+4}\wedge H_3\;.
\)
This is the differential $d_7$ in the twisted AHSS. 
Note, again, that because $H_3$ is closed odd degree form and due to 
\eqref{nextH}
 we have the 
closedness of the Massey quadruple product, i.e.,
 $d \langle H_3, H_3, H_3, F_n \rangle=0$.
We can continue in this fashion until we exhaust the possible degrees allowed by
our dimension, in this case 10. So if we do not trivialize simply by being above dimension 10, then 
we could start with a degree 2 RR field $F_2$ and form a quadruple 
Massey product, leading to $F_6$, and so on. 
\end{example}

The above example could be viewed as the cohomological counterpart to 
the homological arguments for modelling the twisted AHSS, given in 
\cite{MMS}. We now consider cohomological trivializations of the Massey product, i.e., find 
the corresponding potentials.

\begin{example}[Massey potentials for twisted RR fields]
Let $F=F_2 + F_4 + F_6 + F_8$ be the inhomogeneous RR form fields with 
$dF_2=0$ (in the absence of $F_0$, i.e., no cosmological constant), so that 
$F_2$ represents a cohomology class. Even though classically the class
$[F_2]$ is annihilated for dimension reasons 
by the bare differential $Sq^3$ (when working integrally),
the class is still acted upon nontrivially by operations arising from the twist. 
The expressions $G=(d- H_3 \wedge)F=G_3 + G_5 + G_7 + G_9$ 
splits into the expressions
$$
G_3=dF_2=0\;, 
\qquad
G_5=dF_4 - H_3 \wedge F_2\;, 
\qquad
G_7=dF_6 -  H_3 \wedge F_4\;, 
\qquad
 G_9=dF_8 - H_3 \wedge F_6\;.
$$
Then the class $[G_5]=-[H_3]\cup [F_2]$ represents $d_3[F_2]$, so that the 
differential $d_3$ in the tAHSS is just multiplication by $H_3$. If $[H_3 \wedge F_2]=0$,
so that $H_3 \wedge F_2=dF_4$ then this makes $G_5=0$. Then 
$G_7=dF_6 - H_3 \wedge F_4$ represents $d_5 [F_2]$ given by the triple
Massey product 
$$
d_5[F_2]=-\langle H_3 , H_3, F_2 \rangle\;.
$$
Continuing in a similar fashion, we see that 
$$
d_7[F_2]=-\langle H_3 , H_3, H_3, F_2 \rangle 
$$
and so on.

\end{example}

\begin{remark}
{\bf (i)} In the above examples we could have taken our starting point any of the fields
$F_i$. However, we choose to start with the lowest term $F_2$ to illustrate that
all the fields can be accounted for via a physical modelling of   the differentials of  the 
tAHSS. Furthermore, the ring of invariants identified in \cite[Prop. 8.8]{AS2}
 (see \cite{BeM} for an explicit list) will contain  the class $F_2$ in every relevant degree. 
\item {\bf (ii)} The fact that there are no odd (rational) characteristic classes for 
twisted K-theory aside from the twisting class (\cite[Sec. 8]{AS2}) is compatible with 
the fact that the fields in type IIA string theory, classified by $K^0(X; H_3)$, 
are all of even degree. 
\item {\bf (iii)} The above examples have counterparts in type IIB string theory,
where the RR fields are of odd degrees.  Here we start with $F_1$ and generate all the 
other fields similarly. Again, the ring of invariants will involve $F_1$ in all relevant degrees. 
\end{remark} 

\medskip
We now would like to find the relationship between the Massey products on the higher differentials in the spectral 
sequence  for twisted differential K-theory. As in \cite{AS2}, we need to work rationally. For twisted differential K-theory, 
the correct  replacement is twisted differential periodic rational cohomology 
$\widehat{H}_{\hat{h}}^*(X;\QQ[u,u^{-1}])$  (see \cite{GS4}\cite{GS6}), 
where we  regard the twist $\hat{h}:X\to \BB^2U(1)_{\nabla}$ as a twist for periodic rational cohomology via the 
canonical map 
$\BB^2U(1)_{\nabla}\to \BB^2\RR/\QQ_{\nabla}$. \footnote{The latter stack can be presented via the Dold-Kan 
correspondence by the complex 
$$
\QQ\into \Omega^0\overset{d}{\to} \Omega^1\overset{d}{\to} \Omega^2 \to \hdots \;,
$$
i.e., simply replace $\ZZ$ by $\QQ$ in the Deligne complex.} 
In \cite{GS1} we established the basic theory of differential Massey products. Algebraically, these products 
end up behaving exactly as their classical counterparts -- one simply replaces the wedge
 product with the Deligne-Beilinson cup product operation $\cup_{\rm DB}$.

\begin{example}[Differential Massey products] 
Let $\hat{h}:X\to \BB^2\RR/\QQ_{\nabla}$ be a cocycle in (rational) Deligne cohomology refining $H$ and let 
$\hat{x}:X\to \BB^{p-1}\RR/\QQ_{\nabla}$ be a cocycle. Suppose there is $\hat{y}:X\times \Delta[1]\to \BB^{p+2 }\RR/\QQ_{\nabla}$ 
such that $D(\hat{y})=\hat{h}\cup_{\rm DB}\hat{x}$, where $D=d+(-1)^{p+1}\delta$ is the {\v C}ech-Deligne differential. 
By graded commutativity, $2\hat{h}\cup_{\rm DB}\hat{h}=0$ and since we are working over $\QQ$, this implies 
$\hat{h}\cup_{\rm DB}\hat{h}=0$. Then we can form the cochain
\(\label{masdfcoh}
\hat{y}\cup_{\rm DB}\hat{h}:X\times \Delta[1]\xymatrix{\ar[r]&} \BB^{p+5}\RR/\QQ_{\nabla},
\)
representing an element in $\pi_1\map(X,\BB^{p+5}\RR/\QQ_{\nabla})\cong
 \pi_1\map(X,\BB^{p+5}\RR/\QQ^{\delta})\cong H^{p+4}(X;\RR/\QQ)$. The cocycle \eqref{masdfcoh}
  is an element of the Massey product $\langle \hat{h},\hat{h},\hat{x}\rangle$ which necessarily lands 
in the flat part of differential cohomology. Modulo ambiguity in the Massey product, the restriction of this operation to 
$H^{p-1}(X;\RR/\QQ)\into \widehat{H}^p(X;\RR/\QQ)$ gives a map
$$
\langle \hat{h},\hat{h},-\rangle:H^{p-1}(X;\RR/\QQ)\xymatrix{\ar[r]&}  H^{p+4}(X;\RR/\QQ),
$$
raising degree by 5.
\end{example}

This example indicates that the differential Massey products always represent \emph{flat} differential cohomology classes, and 
in fact this is the case \cite{GS1}. Thus, we can always restrict these operations to flat classes (cohomology with either 
$\RR/\ZZ$ or $\RR/\QQ$ coefficients) and these restrictions are the operations appearing as the differentials in the AHSS. 

\medskip
The $E_2$-page of the AHSS for twisted differential K-theory looks identical to the untwisted case with one exception. In the twisted case, we have $E_2^{0,0}=\Omega^{\rm even}_{\ZZ,d_H\text{-}{\rm cl}}(X)$, the group of \emph{twisted} closed forms of even degree with degree zero 
component $\omega_0\in \ZZ$ (see \cite{GS6} for details). The following proposition was proved in \cite{GS3}.

\begin{proposition}[Higher differentials in twisted differential K-theory]\label{thefldfrat}
Let $\widehat{h}:X\to \BB^2U(1)_{\nabla}$ be a twist for differential $K$-theory, regarded as a twist for periodic 
rational cohomology via  the differential Chern  character map (see Prop. \ref{prop-twdch}). Then the differentials 
$d_{2p+1}$ can be identified with the differential  Massey product operation 
$$
d_{2p+1}= -  \langle \; \underbrace{ \hat{h},\hat{h}, \hdots, \hat{h}}_{k~{\rm times}} \; , -\rangle\;.
$$
\end{proposition}

\begin{remark}[Rational vs. non-rational differentials] \label{Q-not}
Non-rationally, there is not much we can say, since these differentials have not been identified even in the topological case.
 In parallel to the topological case, we do however have the identification 
$$
d^{p,-q}_{3}=\widehat{Sq}^3+ \widehat{h}\cup_{{}_{\rm DB}} (-) \;,
$$
for $q>0$, where $\widehat{Sq}^3$ is the again the operation 
$
jSq^2\rho_2\beta
$ as before.
\end{remark}

\medskip
As in the untwisted case, the differentials in the AHSS split into two types (cf. \eqref{type 1 diff} and \eqref{type 2 diff}).
The flat differentials, which we have identified rationally in Prop. \ref{thefldfrat} and differentials of the form 
$$
d_p^{0,0}:E^{0,0}_{p-1}\subset \Omega^{\rm even}_{d_{H}\text{-}{\rm cl}}(X)
\longrightarrow \ker(d_{p-1})\subset H^{p}(X;U(1)).
$$ 
In \cite{GS6} we showed that for $p$ be an even integer, the differential $d_p^{0,0}$ take the form
$$
\xymatrix{
d_p: \Omega^{\rm even}_{d_{H}\text{-}{\rm cl}}(X) \; \ar[r] &
  H^{p}(X;\RR/\QQ)
},
$$
where $\Omega^*_{\QQ,d_{H}\text{-}{\rm cl}}(X)$ is the subgroup of twisted closed forms with degree zero term $\omega_0$
 given by a constant function taking values in $\QQ$. Moreover, the differential $d_p$ maps a twisted closed form of the type
$\omega=0+0+\hdots+\omega_p+\omega_{p+2}+\hdots$
to the class of the leading term $\omega_p$, modulo $\QQ$, i.e.,
\(
\label{ratdfah}
d_p(0+0+\hdots +\omega_p+\omega_{p+2}+\hdots)=[\omega_p]\mod \QQ\;.
\)

\medskip
More generally, for twisted differential cohomology, we find the following.
\begin{proposition}[Lifting flat classes to twisted differential K-theory]\label{necvcon}
A necessary condition for lifting a \emph{flat} differential cohomology class $\hat{x} \in \widehat{H}^{i}(M;\ZZ)$
 is the vanishing of the action 
of the differential in the AHSS on that 
class. That is, 
$$
\widehat{Sq}^3 \hat{x} + \widehat{h}_3 \cup_{{}_{\rm DB}} \hat{x}=0\;.
$$ 
\end{proposition}

Recall that, by definition, $\widehat{Sq}^3=j_2Sq^2\rho_2\beta_{U(1)}$. We can define a differential refinement of 
the 3rd integral Steifel-Whitney class $W_3$ by setting
$$
\xymatrix{
\widehat{W}_3=j_2w_2\in H^2(X;U(1)) \; \ar@{^{(}->}[r]&  \widehat{H}^3(X;\ZZ)
},
$$
which defines a flat differential cohomology class refining $W_3$. For $X$ an oriented 10-manifold and 
$\hat{x}\in \widehat{H}^7(X;\ZZ)$ a flat differential cohomology class, the Wu formula implies that 
$$
\widehat{Sq}^3\hat{x}=j_2Sq^2\beta_{U(1)}x=
j_2w_2\cup \beta_{U(1)}x=\widehat{W}_3\cup_{\rm DB} \hat{x}\;.
$$
Note also that the cup product $\widehat{W}_3\cup_{\rm DB} \hat{x}$ is invariant under the variation 
$\widehat{W}_3\mapsto \widehat{W}_3+\alpha$, with $\alpha\in \mathscr{J}^2(X)=H^2(X;\RR)/H^2(X;\ZZ)$ 
the intermediate Jacobian. Thus, we might as well assume $\widehat{W}_3$ is an arbitrary differential refinement of 
$W_3$ with vanishing curvature. 

\subsection{Anomalies} 
\label{anomalies}

We now explain how  to  refine the Freed-Witten anomaly \cite{FW} to the differential setting and relate to the 
above constructions. Recall that in \cite{FW}, it was shown that the Pfaffian of the Dirac operator on the worldsheet 
of the string $\Sigma\to X$,  with boundary landing on an oriented submanifold $Q\into X$, is in general not 
well-defined as a function but only as a section  of a line bundle on the space of parameters. This line bundle carries 
a natural metric and flat connection, but the holonomy  of this flat connection is in general nontrivial and is equal to 
$\pm 1$, determined by the second Stiefel-Whitney class $w_2(Q)$. 

\medskip
When the $B$-field vanishes, the relevant factors in the worldsheet path integral are 
\(\label{pthinb0}
{\rm pfaff}(D)\cdot {\rm exp}\Big(i\oint_{\partial \Sigma}A\Big),
\)
where $A$ is the $U(1)$-``gauge field" on $Q$. In general, we have the additional contribution of the $B$-field
\(\label{pthinb}
{\rm pfaff}(D)\cdot {\rm exp}\Big(i\oint_{\partial \Sigma}A+i\int_{\Sigma}B \Big).
\)
In \cite{FW} is was argued that $A$ is not a true gauge field in general, as the curvature may not have integral periods. 
In fact, in order to cancel the anomaly from the Pfaffian, it is necessary for $dA=\mathcal{F}$ to have half-integral periods 
\footnote{Note that we are dropping the prefactors $1/2\pi i$ throughout.} so that its exponential in \eqref{pthinb0} is 
allowed to change sign precisely whenever the Pfaffian does. 

\medskip
In the full differential refinement, the $B$-field is modeled not just by a differential form, but by a full $U(1)$-gerbe with 
connection. The existence of the gerbe $\hat{h}_3:X\to \BB^2U(1)_{\nabla}$ allows us to define a twisted differential 
${\rm Spin}^c$-structure, in the sense of \cite{SSS3}, which generalizes the notion of a twisted ${\rm Spin}^c$ structure 
\cite{Do}\cite{Wan}. In particular, for an oriented submanifold $i:Q\into X$ (to be thought of as a D-brane worldvolume), 
the moduli space of such structures on $Q$ can be identified with the space of sections of the pullback
(see \cite{Cech}\cite{cup}\cite{M5}\cite{E8}) 
\(\label{twdfspc}
\xymatrix{
{\rm Tw}_Q(\BB {\rm Spin}^c_{\nabla})\ar[rr]\ar[d] && 
\BB {\rm Spin}^c_{\nabla}\ar[d]^-{\widehat{W}_3}
\\
Q\ar[rr]^-{i^*\hat{h}_3} && \BB^2U(1)_{\nabla}
}
\)
and this pullback depends on a preferred choice of $\widehat{W}_3$, refining $W_3$. Our goal is to show the following
\begin{proposition}[Differential refinement of the Freed-Witten anomaly] 
A choice of closed differential 2-form $\mathcal{F}$ on $Q$ determines a flat $U(1)$-gerbe with connection refining 
$W_3$ on $Q$. Moreover, taking $\mathcal{F}$ the curvature of $A$ and the corresponding refinement of 
$W_3$ in diagram \eqref{twdfspc}, the obstruction to making 
the quantity \eqref{pthinb} well-defined is precisely the existence of a twisted differential ${\rm Spin}^c$ structure. 
More succinctly, we have the refinement of the Freed-Witten anomaly cancellation
$$
\widehat{W}_3+\hat{h}_3=0\;.
$$
\end{proposition}

The original Freed-Witten anomaly cancellation mechanism says that when $W_3+h_3=0$, there is a choice of 
$U(1)$-``gauge field" on $Q$ for which the potentially anomalous term \eqref{pthinb} is well-defined, but if one is 
given such a field \emph{a-priori}, this choice may not agree with the given field. Thus the difference between our 
anomaly cancellation and the original Freed-Witten anomaly is the specific choice of field $\mathcal{F}$, which 
should be identified with the curvature of the $U(1)$-``gauge field". 

\medskip
Note that every 2-form defines a flat refinement of $W_3$. Indeed, the group of differential refinements of a topological 
torsion class in $H^3(Q;\ZZ)$ with vanishing curvature is a torsor for $H^2(Q;\RR)/H^2(Q;\ZZ)$. Let $w_2(Q)$ 
denote the second Stiefel -Whitney class. Since 
$$
\beta_{U(1)}j_2w_2(Q)=\beta_2w_2(Q)=W_3(Q)\;,
$$
it follows that $j_2w_2(Q)$ defines a refinement of $W_3(Q)$. Hence if $j:H^2(Q;U(1))\into \widehat{H}^3(Q;\ZZ)$
 is the canonical map identifying flat classes with $U(1)$-cohomology classes, we see that every refinement of $W_3$ 
 can indeed be written 
\(
\label{fwanomc}
\widehat{W}_3(Q)=j_2w_2(Q)-j{\rm exp}(\mathcal{F})\;.
\)
Now, for simplicity, let us first consider the case $\hat{h}_3=0$ so that such a structure reduces to a differential 
${\rm Spin}^c$-structure. This, in particular, defines a topological ${\rm Spin}^c$ structure and hence 
we can consider the canonical line bundle $\mathcal{L}$ associated to the ${\rm Spin}^c$ structure. The vanishing 
of the class \eqref{fwanomc} means that the mod 2 reduction of $c_1(\mathcal{L})$ can be obtained through the 
exponential of a 2-form $\mathcal{F}$, which necessarily has half integral periods (i.e. its exponential lands in 
$\ZZ/2\into U(1)$). If we let $\mathcal{F}$ be the curvature of the $A$-field, then this indeed implies that the 
sign ambiguity in   \eqref{pthinb0} is cancelled by the ambiguity in the holonomy, arising from $\mathcal{F}$ 
having only half integral periods. 

\medskip
More geometrically, the cancellation of \eqref{pthinb0} can be obtained from a choice of connection on $\mathcal{L}$. 
Indeed, such a connection determines a differential refinement $\hat{c}_1(\mathcal{L})\in \widehat{H}^2(Q; \Z)$ of 
$c_1(\mathcal{L})$. From the commutative diagram 
$$
\xymatrix{
\widehat{c}_1 \ar@{|->}[d] &
\widehat{H}^2(Q;\ZZ)\ar[r]^-{\frac{1}{2}\mathcal{R}}\ar[d]^-{\mathcal{I}} & \Omega^2_{\rm cl}(Q)\ar[d]
& \mathcal{F} \ar@{|->}[d]
\\
{c}_1 \ar@{|->}[d] &
H^2(Q;\ZZ)\ar[r]^{\frac{1}{2}\times}\ar[d]^-{\rho_2} & H^2(Q;\RR)\ar[d]^-{\rm exp}
& [\mathcal{F}]_\R \ar@{|->}[d]
\\
w_2 &
H^2(Q;\ZZ/2)\ar[r]^{j_2} & H^2(Q;U(1))
& [\mathcal{F}]_{U(1)}
}
$$
and the identify $\rho_2c_1(\mathcal{L})=w_2(Q)$, one sees immediately that if $\mathcal{F}$ is the curvature of $\mathcal{L}$, 
then taking $\tfrac{1}{2}\mathcal{F}$ in \eqref{fwanomc} implies $\widehat{W}_3=0$. In this case, the differential ${\rm Spin}^c$ 
structure is defined completely by a choice of connection on $\mathcal{L}$. In the more general case when $\hat{h}_3\neq0$, 
identifying $B$ with the connection of the gerbe defined by $\hat{h}_3$ shows that the condition 
$$
\widehat{W}_3+\hat{h}_3=0
$$
is precisely what is needed to make \eqref{pthinb} well-defined.

\begin{remark}
[A consequence of differential refinement] 
Another advantage of the full refinement of the Freed-Witten anomaly is that it gives a precise geometric meaning to the $A$-field, 
even when it cannot be globally identified with a $U(1)$-gauge field. It can be identified with a choice of differential refinement of 
$W_3$ with vanishing curvature. 
\end{remark}

\section{Explicit classification of RR fields in traditional backgrounds}
\label{Sec-comp}

\subsection{Twisted differential K-theory of spheres} 
\label{Sec-sph}

Spheres are important compactification spaces for string theory when considering background fluxes. 
We have already considered integrality conditions for (even) spheres and projective spaces
in Example \ref{Ex-S2n} and Example \ref{Ex-cpn}. 
Here we consider RR fields on the 3-sphere $S^3$, using the careful calculations in \cite{GS5}, 
which generalize to differential twisted K-theory 
the corresponding twisted K-theory calculations of the Lie group $\op{SU}(2)$
\cite{MMS}\cite{BCMMS}\cite{Braun}\cite{Do}\cite{FHT}\cite{MR}\cite{Ro2}. 
This was also studied in \cite{CMW} using index and group theoretic methods. 

\medskip
Let $h:S^3\to K(\ZZ,3)$ be a map representing an element  (which we also denote by $h$) in integral 
cohomology $H^3(S^3;\ZZ)\cong \ZZ$ and denote the corresponding twisted $K$-theory  on $S^3$ by ${K}^*_h(S^3)$. 
Note that the map $h$ can be refined to a gerbe with connection $\widehat{h}:S^3\to \BB^2U(1)_{\nabla}$, whose curvature form 
is $H$.  Now we consider $(\Omega^*[u,u^{-1}],d_{H})$, the sheaf of periodic, $H$-twisted de Rham complex on $S^3$, with differential 
$d+H\wedge (-)$.  Thus, the triple $\widehat{\mathscr{K}}_{{\hat h}}:=(\mathscr{K}_{h}, {\rm ch}, (\Omega^*[u,u^{-1}],d_{H}))$ 
gives the data of a differential refinement of the $h$-twisted $K$-theory 
spectrum, and we denote the underlying theory by $\widehat{K}_{\hat{h}}^*(S^3)$.

\medskip
The calculations via the Mayer-Vietoris sequence or the twisted differential AHSS \cite{GS5} 
give the following. 

\begin{example}
[RR fields in type IIA on the 3-sphere]
Let $\hat{h}:S^3\to \BB^2U(1)_{\nabla}$ be a differential twist as a gerbe with connection. 
Recall (see Remark \ref{Q-not})  that we identified the differential on the $E_3$-page as 
$d_3=\widehat{Sq}^3_{\ZZ}+\hat{h}\cup_{\rm DB}$. 
For the 3-sphere, the $U(1)$-cohomology is calculated from the exponential sequence as 
$$
H^2(S^3;U(1))\cong 0 \qquad \text{and} \qquad 
 H^3(S^3;U(1))\cong U(1) .
$$
Then for $\widehat{K}^0$ we see that all relevant differentials must vanish 
and the spectral sequence collapses at the $E_2$-page in Diagram 
\eqref{E2-Kth}. There is no extension problem in this case, and we have the isomorphism
$$
\widehat{K}^0(S^3; {\hat{h}})\cong (\Omega^{\rm even}(S^3),d_{H})_{\rm cl}\cong\Omega^2(S^3)_{\rm cl} .
$$
This means that the RR fields in type IIA string theory on $S^3$ are classified by closed 2-forms on 
the 3-sphere, an instance of which would be a flat abelian 2-gerbe connection. 
\end{example}

\begin{example}
[RR fields in type IIB on the 3-sphere]
For $\widehat{K}^1$, we need to calculate the kernel of the differential $d_3$ as the map
$$
d_3:U(1)\longrightarrow U(1)\cong H^3(S^3;U(1)) .
$$
For degree reasons, the refined integral Steenrod square 
$\widehat{Sq}^3_{\ZZ}$ vanishes on $U(1)$, which reduces the task to finding the kernel of $\hat{h}$. 
The formula for the Deligne-Beilinson cup product $\cup_{\rm DB}$ sends an element in $U(1)$, written in complex form 
as $e^{2\pi i \theta}$, to the element $e^{2\pi i h\theta}$, where $h$ is the integer representing the underlying topological twist. 
Hence the kernel can be identified with the $h$-roots of unity, which as an abelian group is isomorphic to $\ZZ/h\ZZ$. 
Furthermore, for degree reasons, there are no nontrivial differentials out of the term $(\Omega^*(S^3),d_{H})_{\rm cl}$. In this case,
 there is no extension problem and we arrive at the isomorphism
$$
\widehat{K}^1(S^3; {\hat{h}})\cong \ZZ/h\ZZ\oplus (\Omega^*(S^3),d_{H})_{\rm cl} .
$$
In particular, this identification shows that every twisted closed odd RR form lifts to $\widehat{K}^1$. 
\end{example}

\subsection{The (twisted) differential K-theory of tori}
\label{Sec-tori}

Tori play an important role in (flat) compactifications of string theory. 
We begin with some preliminary computations which describe special instances of the RR fields. 

\begin{lemma} [RR fields on the $k$-torus $T^k$] The K-theory of the $T^k$ is given by 
$$K^0(T^k)=\bigoplus_n \Lambda_{\ZZ}^{2n}(x_1,x_2,\hdots,x_k)
\qquad 
\text{and}
\qquad K^1(T^k)=\bigoplus_n \Lambda_{\ZZ}^{2n+1}(x_1,x_2,\hdots,x_k)
$$
where the exterior algebras are taken over $\ZZ$.
\end{lemma}
\theproof
The cohomology of the $k$-torus is given by the exterior algebra
$$
H^*(T^k;\ZZ)\cong \Lambda^*(x_1,x_2,\hdots,x_k),
$$
where $x_i$ are generators of $H^1(T^k)\cong \ZZ^k$. Since the cohomology groups contain no torsion, 
the AHSS degenerates at the $E_2$-page and there is no extension problem. This immediately implies the claim.
\endofproof

Note that the above also follows from applying the general results of Hodgkin \cite{Hod} on the K-theory of Lie groups.  
The following is then a direct consequence of the long exact sequence for K-theory with coefficients induced by the 
exponential sequence.

\begin{corollary}[Flat RR fields on  the $k$-torus] 
\label{k-thu1}
The K-theory with $U(1)$-cofficients of the $T^k$ is given by 
\begin{align*}
K_{U(1)}^0(T^k) &=\bigoplus_n \Lambda_{\RR}^{2n}(x_1,x_2,\hdots,x_k)/\Lambda_{\ZZ}^{2n}(x_1,x_2,\hdots,x_k)
\cong \bigoplus_n\mathscr{J}^{2n}(T^k),
\\
K_{U(1)}^1(T^k)&=\bigoplus_n \Lambda_{\RR}^{2n+1}(x_1,x_2,\hdots,x_k)/\Lambda_{\ZZ}^{2n+1}(x_1,x_2,\hdots,x_k)
\cong \bigoplus_n\mathscr{J}^{2n+1}(T^k),
\end{align*} 
where $\mathscr{J}^{m}(T^k)$ is the intermediate Jacobian $H^{m}(T^k;\RR)/H^m(T^k;\ZZ)$. 
\end{corollary}

\begin{proposition}[Geometric RR fields on the $k$-torus] 
The differential K-theory of the $k$-torus is given by 
$$
\widehat{K}^0(T^k)\cong \bigoplus_{n} \mathscr{J}^{2n+1}(T^k)\oplus \Lambda_{\ZZ}^{2n}
(\omega_1,\omega_2,\hdots,\omega_k)\oplus d\Omega^{2n+1}
$$
where $\omega_i$ are normalized harmonic form representatives for the generators of $H^1(T^k;\ZZ)$. Similarly,
$$
\widehat{K}^1(T^k)\cong \bigoplus_{n} \mathscr{J}^{2n}(T^k)\oplus 
\Lambda_{\ZZ}^{2n+1}(\omega_1,\omega_2,\hdots,\omega_k)\oplus d\Omega^{2n}.
$$ 
This isomorphism identifies $\omega_i\wedge \omega_j$ with a geometric representative for the first Chern class 
and $\theta\omega_i\in \mathscr{J}^{1}(T^k)$ with the Chern-Simons invariants of flat bundles.
\end{proposition}
\theproof
Since $T^k$ is a formal manifold, there is a choice of metric so that the Hodge decomposition gives rise to an identification 
$\Omega^{\rm even}_{\rm cl}(T^k)\overset{g}{\cong} \Lambda_{\RR}(\omega_1,\hdots,\omega_k)\oplus d\Omega^{\rm odd}(T^k)$,
where $\omega_i$ are the unique harmonic forms representing the generator $1\in H_{\rm dR}^1(S^1)\cong \RR$. From the identification 
in Corollary \ref{k-thu1}, the AHSS for $\widehat{K}$ degenerates at the $E_2$-page and the only condition on forms that they lift to 
$\widehat{K}$ is that they have integral periods. This shows that $\widehat{K}^0(T^k)$ fits into an exact sequence
$$
0\longrightarrow \bigoplus_n\mathscr{J}^{2n+1}(T^k)\longrightarrow 
\widehat{K}^0(T^k)\longrightarrow
  \Lambda_{\ZZ}(\omega_1,\hdots,\omega_k)\oplus d\Omega^{\rm odd}(T^k).
$$
Since $\mathscr{J}^{2n+1}(T^k)$ is a divisible group, this sequence splits. The claim for $\widehat{K}^1$ is 
proved similarly.
\endofproof

\begin{remark}[Geometric RR fields with background NS-field and twisted differential K-theory]
We observe that for a differential refinement $\widehat{W}_3$ of $W_3$, we necessarily have 
$\widehat{W}_3=j{\rm exp}(\mathcal{F})$, for some closed 2-form $\mathcal{F}$. If $\mathcal{F}$ 
has integral periods, then $\widehat{W}_3=0$ and the twisted differential K-theory reduces to the untwisted.
\end{remark}

\subsection{The (twisted) differential K-theory of Calabi-Yau threefolds} 
\label{Sec-CY} 

Now we consider Calabi-Yau manifolds, a third main class of compactification spaces for type II
string theory. In \cite{DM} it was shown that for a Calabi-Yau threefold $M$ one has the identification
\(
\label{intkid2}
\widetilde{K}^0(M)\cong H^2(M;\ZZ)\oplus H^4(M;\ZZ)\oplus 2\cdot H^6(M;\ZZ) ,
\)
where the isomorphism is exhibited by taking the first, second and third Chern class \cite{DM}. 
The method of proof  is direct and amounts to a careful consideration of the $7$th stage of the Postnikov tower for 
the classifying space $B{\rm SU}$. Alternatively, one can compute these groups (modulo extension) via the AHSS. 
If one presents the AHSS using the filtration on the $K$-theory spectrum via its Postnikov stages, then solving the
extension problem amounts to the same consideration in \cite{DM}.

 \medskip
 In preparation for our computation of differential cohomology, we prove the following.
 
\begin{lemma} [Flat RR fields and flat K-theory of 6-manifolds] \label{lemu1k}
Let $M$ be a closed oriented 6-dimensional manifold.
\item {\bf (i)} We have an isomorphism 
$$
\widetilde{K}_{U(1)}^{-1}(M)\cong {\rm Tor}(\widetilde{K}^0(M))\oplus 
\bigoplus_{i=1}^3\mathscr{J}^{2i-1}(M) ,
$$
where $\mathscr{J}^{2i-1}(M)=H^{2i-1}(M;\RR)/H^{2i-1}(M;\ZZ)$ is the intermediate Jacobian 
and $\widetilde{K}^0(M)$ is computed as in \eqref{intkid}. 
\item {\bf (ii)} If $M$ is a Calabi-Yau threefold, then we further have 
$$
\widetilde{K}^{-1}_{U(1)}(M)\cong \bigoplus_{i=1}^3{\rm Tor}(H^{2i}(M;\ZZ))\oplus 
\bigoplus_{i=1}^3\mathscr{J}^{2i-1}(M)\cong \bigoplus_{i=1}^3H^{2i-1}(M;U(1)).
$$
\end{lemma}
\theproof
From the Bockstein sequence, we have an exact sequence
$$
\widetilde{K}^{-1}_{\RR}(M)\longrightarrow
 \widetilde{K}^{-1}_{U(1)}(M)\longrightarrow
  \widetilde{K}^0(M)\longrightarrow \widetilde{K}_{\RR}(M) .
$$
This gives an exact sequence $\widetilde{K}^{-1}_{\RR}(M)\to  \widetilde{K}^{-1}_{U(1)}(M)\to {\rm Tor}(\widetilde{K}^0(M))\to 0$, 
which immediately implies the first result. For the second, the Wu formula implies that $Sq^2:H^4(M;\ZZ/2)\to H^6(M;\ZZ/2)$ is 
representable by cup product with $w_2$. Since any Calabi-Yau is ${\rm Spin}^c$ and $c_1=0$, it follows that $0=rc_1=w_2$ 
and $Sq^2$ vanishes. Therefore, ${\rm Tor}(\widetilde{K}^0(M))\cong \bigoplus_{i=1}^3{\rm Tor}(H^{2i}(M;\ZZ))$ by  
\eqref{intkid2}.  The final identification follows from the (noncanonical) decomposition
$$
H^k(M;U(1))\cong H^k(M;\RR)/H^k(M;\ZZ)\oplus {\rm Tor}(H^{k+1}(M;\ZZ))\;.
$$

\vspace{-.7cm} 
\endofproof

We now consider the fully differential case. 

\begin{proposition}  [Geometric RR fields and differential K-theory of 6-manifolds] 
\label{dfkcaly}

\item {\bf (i)} The differential $K$-theory of a compact oriented 6-dimensional manifold fits into an exact sequence
\(\label{extpr1}
{\rm Tor}(\widetilde{K}^0(M))\oplus \bigoplus_{i=1}^3\mathscr{J}^{2i-1}(M)\longrightarrow
 \widehat{K}^0(M) \longrightarrow {\rm Im}({\rm ch}) ,
\)
where $\mathscr{J}^{2i-1}(M)\cong H^{2i-1}(M;\RR)/H^{2i-1}(M;\ZZ)$ is the intermediate Jacobian. 
Moreover, this sequence splits, but not canonically.

\item {\bf (ii)} Let $e: H^*(M; \R) \to H^*(M; U(1))$ denote the exponential map arising from 
coefficients. The image of the Chern character is given by 
$$\hspace{-.1cm}
{\rm Im}(\ch) \!\!=\!\!\Big\{\!(\ch_1, \ch_2, \ch_3)\in \bigoplus_{i=1}^3\Omega_{\rm cl}^{2i}(M)\ | \; \ch_1\in H^2(M;\ZZ),  e(\ch_2)=j_2(\overline{c}_1^2),e(\ch_3)=j_2(\overline{Sq}^2(\ch_2))-j_3(\overline{c}^3_1) \!\Big\},
$$
where $(c_1,c_2,c_3)$ denote the first, second and third Chern class.


\item {\bf (iii)} For $M$ a Calabi-Yau threefold, we have
$$
\widehat{K}^0(M)\cong \bigoplus_{i=1}^3{\rm Tor}(H^{2i}(M;\ZZ))\oplus \bigoplus_{i=1}^3
\mathscr{J}^{2i-1}(M)\oplus {\rm Im}({\rm ch}) ,
$$
with
$$
{\rm Im}(\ch)=\Big\{(\ch_1, \ch_2, \ch_3)\in \bigoplus_{i=0}^3\Omega_{\rm cl}^{2i}(M)\ | \ 
e(\ch_2)=j_2\overline{c}_1^2, e(\ch_3)=-j_3\overline{c}_1^3\Big\}.
$$
The isomorphism identifies $\ch_1, \ch_2, \ch_3,c_1,c_2$ and $c_3$ with the 
Chern characters forms and Chern classes, respectively. The torsion group is identified with torsion Chern 
classes and the intermediate Jacobian is identified with Chern-Simons classes.
\end{proposition}
\theproof
For a 6-dimensional manifold, the fact that we have such an exact sequence follows from Lemma \ref{lemu1k} and from the diagonal 
sequence in the differential cohomology diamond \eqref{kodfdiam}. It remains to calculate the image of the Chern character. For this, 
we apply the AHSS for differential $K$-theory developed in \cite{GS3}\cite{GS5}. The extension in \eqref{extpr1} is precisely the 
final extension problem for the refined AHSS corresponding to the filtration level
$F_0\widehat{K}(M):=\ker(i^*_0:K(M)\to K(F_0(\check{C}(\{U_{\alpha}\}))$, where 
$F_0$ denotes the 0th level of the filtration. Thus the permanent cycles in 
$$
\ZZ\oplus \bigoplus_{p=1}^{3}\Omega^{2p}(M)=E^{0,0}_2 \; \Longrightarrow \; E^{0,0}_{\infty}
$$ 
compute the image of ${\rm ch}$. Because of the low dimensions, the only possible nonvanishing odd differentials are 
(see Proposition \ref{dfrefbu}) 
\(\label{fldfident}
 d_3=j_2Sq^2\rho_2\beta_{U(1)} 
 \qquad \text{and} \qquad 
  d_5=j_2Sq^2\beta_{U(1)}+j_2\mathscr{P}^1_3\beta_{U(1)} ,
\)
while the even differentials are given by (see Propositions \ref{iddifd2}, \ref{d4iddf}, and \ref{prop-deg6})
\(\label{crdfident}
 d_2={\rm exp}, \quad d_4={\rm exp}+j_2Sq^2, \quad \text{and} \quad  d_6={\rm exp}+j_2\overline{Sq}^2+j_3\mathscr{P}^1_3\;.
\)
From these identifications, we find that the permanent cycles in $E^{0,0}_{\infty}$, which is the image of the Chern 
character, are those forms such that
\begin{align*}
[\ch_1] \!\!\!\!\mod \ZZ & =0 ,
\\
[\ch_2] \!\!\!\!\mod \ZZ &\equiv  j_2\overline{c}_1^2 ,
\\
[\ch_3] \!\!\!\!\mod \ZZ &\equiv  j_2\overline{Sq}^2r(\ch_2)- j_3\overline{c}_1^3 .
\end{align*}
This implies the first claim. For the splitting, fix a metric $g$ on $M$ and 
consider the corresponding Hodge decomposition on forms. The group ${\rm Im}(\ch)$ is free abelian and defines a maximal rank 
lattice in $\bigoplus_iH^{2i}(M;\RR)\cong \bigoplus_i{\rm harm}^{2i}(M)$. From the basic properties of ${\rm Ext}$, one then 
sees that the splitting will follow provided ${\rm Ext}^1(d\Omega^k(M),\ZZ/n)=0$ for any integers $n,k$. This is easily deduced 
via the injective resolution $\ZZ/n\into U(1)\overset{\times n}{\to} U(1)$ and the commutative diagram 
$$
\xymatrix{
d\Omega^k(M)\ar[rr]^{\times (1/n)}\ar[d]^-{\phi} && d\Omega^k(M)\ar[d]^-{\phi}
\\
U(1)&&\ar[ll]_-{\times n} U(1)
\;.
}
$$
For the 
Calabi-Yau case, the simplification of ${\rm Im}(\ch)$ follows from the Wu formula and vanishing of $c_1$ 
(i.e., $\overline{Sq}^2(\ch_2)=0$). 
\endofproof

This immediately implies the following divisibility conditions and congruences. 

\begin{corollary}[Periods for ${\rm CY}_3$]
For a Calabi-Yau threefold $M$ and any vector bundle $E\to M$, $\ch_3(E)$ has periods in $\frac{1}{3}\ZZ$, 
i.e., $c_3$, $c_1c_2$ and $c_1^3$ are all divisible by 2. For the tangent bundle, $\ch_1$, $\ch_2$ and $\ch_3$ 
are all integral.
\end{corollary}

The above is also useful, for instance, in interpreting the Chern character as an integral twist, e.g. of 
a String structure in the context of the Green-Schwarz anomaly cancellation in the heterotic 
case \cite{SSS3}. 

\medskip
We now turn our attention to the computation of $\widehat{K}^1$ for the type IIB case. 
In \cite{DM}, it was also  shown that for a Calabi-Yau  threefold $M$ one has
\(
\label{intkid}
\widetilde{K}^1(M)\cong H^1(M;\ZZ)\oplus H^3(M;\ZZ)\oplus H^5(M;\ZZ) .
\)
From this identification, \eqref{intkid2} and the exponential sequence for $K$-theory with coefficients, we immediately find that 
\(
\widetilde{K}_{U(1)}^{-2}(M)\cong \bigoplus_{i=0}^3{\rm Tor}(H^{2k}(M;\ZZ))\oplus \mathscr{J}^1(M)\oplus 
\mathscr{J}^4(M)\oplus \tfrac{1}{2}\mathscr{J}^6(M) ,
\)
where $ \frac{1}{2}\mathscr{J}^6(M)\cong H^6(M;\RR)/2\cdot H^6(M;\ZZ)$. Application of the AHSS then gives the following 

\begin{proposition} [Geometric RR fields in Type IIB on a Calabi-Yau threefold] 
\label{prop-iib}
We have an identification 
$$
\widehat{K}^1(M)\cong \bigoplus_{i=1}^3{\rm Tor}(H^{2i-1}(M;\ZZ))\oplus 
\mathscr{J}^1(M)\oplus \mathscr{J}^4(M)\oplus \tfrac{1}{2}\mathscr{J}^6(M)\oplus
 \bigoplus_{i=1}^3\Omega_{{\rm cl},\ZZ}^{2i-1}(M) ,
$$
where $\Omega_{{\rm cl},\ZZ}^{2i-1}(M)$ is the group of closed $(2i-1)$-forms with integral periods. The isomorphism 
identifies the forms and torsion part with the odd Chern classes (i.e. the generators of 
$H^*(U(n);\ZZ)\cong \Lambda^*(a_1,a_3,\hdots, a_{2n-1})$) and the intermediate Jacobians with Chern-Simons classes 
associated with the odd characteristic forms. 
\end{proposition}
\theproof
From the identification of the differentials as in Proposition \ref{dfkcaly} we find that, from the Wu formula and by 
degree considerations, all differentials vanish and the spectral sequence collapses. 
\endofproof

The results of Propositions \ref{dfkcaly} and \ref{prop-iib} exhibit
 the richness of  describing the RR fields  by twisted differential K-theory in a Calabi-Yau background, 
which amounts to specifying the following data
\begin{enumerate}
\vspace{-2mm} 
\item Purely topological: torsion Chern classes, given by the Tor term. 
\vspace{-2mm} 
\item Purely geometric: Chern character forms, given by the last factor.
\vspace{-2mm} 
\item Mixed data: the intermediate Jacobians as the Chern-Simons invariants with values in $U(1)$. 
\end{enumerate}

We end by considering a special case of the twisted setting for both type IIA and type IIB fields, i.e., for 
both $\widehat{K}^0$ and $\widehat{K}^1$.

\begin{remark}[Twist by the differential third Stiefel-Whitney class]
Let $\widehat{W}_3$ be a refinement of $W_3$, determined by a 2-form $\mathcal{F}$ as in Section \ref{anomalies}. For 
the twist $\widehat{W}_3=\hat{h}_3$, we have $\widehat{W}_3=j{\rm exp}\big(\mathcal{F}\big)$. This follows from 
the fact that any Calabi-Yau is spinnable and hence $j_2w_2=0$. If $\mathcal{F}$ has integral periods, then the twist 
vanishes and the $\widehat{W}_3$-twisted differential K-theory is isomorphic to the 
underlying untwisted differential theory.
\end{remark}


\end{document}